\documentclass[pdftex,twocolumn,epjc3]{svjour3}          % twocolumn
\RequirePackage[T1]{fontenc}
\smartqed  % flush right qed marks, e.g. at end of proof
\RequirePackage{graphicx}
\RequirePackage{mathptmx}      % use Times fonts if available on your TeX system
\RequirePackage{flushend}
\RequirePackage[numbers,sort&compress]{natbib}
\RequirePackage[colorlinks,citecolor=blue,urlcolor=blue,linkcolor=blue]{hyperref}
\RequirePackage{amsmath,amssymb}
\RequirePackage{multirow}
\RequirePackage{xspace}
\allowdisplaybreaks

\journalname{Eur. Phys. J. C}

\newcommand{\MET}{\ensuremath{E_\text{T}^\text{miss}}\xspace}

\begin{document}

\title{Phenomenological signatures of additional scalar bosons at the LHC}

\author{Stefan von Buddenbrock\thanksref{e1,addr1}
             \and
             Nabarun Chakrabarty\thanksref{e2,addr2}
             \and
             Alan S. Cornell\thanksref{e3,addr1}
             \and
             Deepak Kar\thanksref{e4,addr1}
             \and
             Mukesh Kumar\thanksref{e5,addr3}
             \and
             Tanumoy Mandal\thanksref{e6,addr4}
             \and
             Bruce Mellado\thanksref{e7,addr1}
             \and
             Biswarup Mukhopadhyaya\thanksref{e8,addr2}
             \and
             Robert G. Reed\thanksref{e9,addr1}
             and
             Xifeng Ruan.\thanksref{e10,addr1}
}

%\thankstext[$\star$]{t1}{Thanks to the title}
\thankstext{e1}{e-mail: stef.von.b@cern.ch}
\thankstext{e2}{e-mail: nabarunc@hri.res.in}
\thankstext{e3}{e-mail: alan.cornell@wits.ac.za}
\thankstext{e4}{e-mail: deepak.kar@cern.ch}
\thankstext{e5}{e-mail: mukesh.kumar@cern.ch}
\thankstext{e6}{e-mail: tanumoy.mandal@physics.uu.se}
\thankstext{e7}{e-mail: bruce.mellado@wits.ac.za}
\thankstext{e8}{e-mail: biswarup@hri.res.in}
\thankstext{e9}{e-mail: robert.reed@cern.ch}
\thankstext{e10}{e-mail: xifeng.ruan@cern.ch}

\institute{School of Physics, University of the Witwatersrand, Johannesburg, Wits 2050, South Africa.\label{addr1}
          \and
          Regional Centre for Accelerator-based Particle Physics, Harish-Chandra Research Institute, Chhatnag Road, 
          Jhunsi, Allahabad - 211 019, India.\label{addr2}
          \and
           National Institute for Theoretical Physics, School of Physics and Mandelstam Institute for Theoretical Physics, 
           University of the Witwatersrand, Johannesburg, Wits 2050, South Africa.\label{addr3}
           \and
           Department of Physics and Astronomy, Uppsala University, Box 516, SE-751 20 Uppsala, Sweden.\label{addr4}
}

\date{Received: date / Accepted: date}

\maketitle

\begin{abstract}
We investigate the search prospects for new scalars beyond the Standard Model (SM) at the Large Hadron Collider (LHC). In these studies two real scalars $S$ and $\chi$ have been introduced in a two Higgs doublet model (2HDM), where $S$ is a portal to dark matter (DM) through its interaction with $\chi$, a DM candidate and a possible source of missing transverse energy (\MET). Previous studies focused on a heavy scalar $H$ decay mode $H \to h\chi\chi$, which was studied using an effective theory in order to explain a distortion in the Higgs boson ($h$) transverse momentum spectrum~\cite{vonBuddenbrock:2015ema}. In this work, the effective decay is understood more deeply by including a mediator $S$, and the focus is changed to $H \to h S,~SS$ with $S \to \chi\chi$. Phenomenological signatures of all the new scalars in the proposed 2HDM are discussed in the energy regime of the LHC, and their mass bounds have been set accordingly. Additionally, we have performed several analyses with final states including leptons and \MET, with $H \to 4 W$, $t(t)H \to 6 W$ and $A \to ZH$ channels, in order to understand the impact these scalars have on current searches.

%With $m_h = 125$~GeV, we propose the masses of other scalars as $2 m_h < m_H < 2 m_t$ for the heavy Higgs $H$, $m_A > m_H + m_{Z}$ for the pseudoscalar $A$, $m_H + m_{Z} < m_{H^\pm} < m_A$ for the charged Higgs bosons, $m_h < m_S < m_H - m_h$ and  $m_\chi \sim m_h/2$ for our studies.
\end{abstract}

\section{Introduction}
\label{intro}
In light of the discovery of a Higgs-like scalar~\cite{Englert:1964et,Higgs:1964ia,Higgs:1964pj,Guralnik:1964eu,ATLAS:2012gk,CMS:2012gu} at the Large Hadron Collider (LHC), there have been many studies devoted to understanding the scalar's properties and couplings to Standard Model (SM) particles. In general, two lines of investigation have been pursued: (a) experimental analyses to closely examine if the behaviour of this scalar reveal any discrepancy with predictions of the SM, and 
(b) theoretical studies on how any new physics -- both model-dependent and independent -- can be discerned.  
The `new physics' possibilities in this context often stress the possible presence of additional scalars which may
participate in electroweak symmetry breaking (EWSB). As such, searches for new scalars, 
neutral and/or charged, are continuously being carried out in various channels by both the ATLAS and CMS collaborations.

There are many possible theoretical models which contain additional scalars. Some of the simplest 
such models are the two-Higgs doublet models (2HDMs)~\cite{gunion, Branco:2011iw}. However, there are a range of issues with these models, such as the generation of neutrino masses that can accommodate a 125~GeV scalar, especially for supersymmetric models~\cite{King:2015aea}. This includes Higgs-like scalars belonging to representations of SU(2), which are not necessarily doublets. 
Furthermore, the source of dark matter (DM) in the universe remains unresolved, and many hypotheses have been put forward in an attempt to explain its origin and existence~\cite{Abdallah:2015ter}.

If any new physics exists in the scalar sector (especially within the reach of the LHC) it should be observed by the experimental collaborations in the near future. With this in view, possible sources of deviation from the SM could be inferred by looking at fiducial Higgs production cross sections and differential distributions~\cite{Aad:2015lha,Aad:2016lvc,Khachatryan:2015yvw,Khachatryan:2015rxa,Khachatryan:2016vnn}. Several of the distributions in this area of study -- most notably the Higgs boson transverse momentum ($p_\text{T}$) spectrum -- are sensitive to new physics predictions, and it is an interesting study to identify if new physics models can provide a compatible description of the data.

The present work is an effort in this direction, where we shall study the model-dependence and independence of a Type-II inspired 2HDM. 
Our addition to the standard 2HDM shall be to include a singlet scalar, $\chi$, which is made odd under 
a ${\mathbb Z}_2$ symmetry (and is thus stable for qualification as a DM candidate). In a previous study~\cite{vonBuddenbrock:2015ema}, the heavier $CP$-even neutral scalar $H$ was assumed to have a large branching ratio (BR) in the channel $H \to
h\chi\chi$ (where $h$ is the 125~GeV Higgs boson) in order to fit the data. This can be facilitated through the on-shell
participation of our additional scalar $S$ in the decay of $H$. The transformation from the effective vertex approach to the $S$ mediated approach can be seen in Fig.~\ref{fig:i} -- this is detailed in Sect.~\ref{theory}. The terms in the Lagrangian involving $\chi$ and $S$ have been included here as effective interaction terms in addition to the Lagrangian of a Type-II 2HDM~\cite{Kumar:2016vut}.

The paper is organised as follows. In Sect.~\ref{theory} we discuss a 2HDM inspired formalism in brief, and then discuss an effective model in Sect.~\ref{hpteff}, by which the Higgs boson $p_\text{T}$ spectrum can be studied. Phenomenological signatures of the new scalars and particles are analysed in Sects.~\ref{pheno} and \ref{anres}. Our findings are then summarised and discussed in Sect.~\ref{conc}.

\section{Framework}
\label{theory}
In this section we briefly discuss the 2HDM with its basic particle content, which we then extend to a Type-II 2HDM. For a more recent review of the constraints in detail, we refer the reader
to Ref.~\cite{Branco:2011iw}. We then introduce two real scalars in this particular Type-II 2HDM, $\chi$ and $S$,
where $\chi$ will be treated as a DM candidate, while $S$ is similar to the SM Higgs boson.

The complete Lagrangian for a 2HDM can be
written as
\begin{align}
{\cal L}_{\text{2HDM}} =& \left( D_\mu \Phi_1\right)^\dag \left( D_\mu \Phi_1\right) + \left( D_\mu \Phi_2\right)^\dag \left( D_\mu \Phi_2\right) \notag \\
&\, - {\cal V}\left(\Phi_1, \Phi_2\right) + {\cal L}_{int},
\label{lag2hdm}
\end{align}
where $\Phi_1$ and $\Phi_2$ are two complex $SU(2)_L$ doublet scalar fields. ${\cal L}_{int}$ contains all possible interaction
terms, including the SM Lagrangian. ${\cal V}\left(\Phi_1, \Phi_2\right)$ is the most 
general remormalisable scalar potential of the 2HDM and may be written as:
 \begin{align}
 {\cal V}\left(\Phi_1, \Phi_2\right) = &\,\, m_1^2 \Phi_1^\dag \Phi_1 + m_2^2 \Phi_2^\dag \Phi_2 
 - m_{12}^2 \left(\Phi_1^\dag \Phi_2 + \text{h.c.} \right) \notag\\
 +& \frac{1}{2} \lambda_1 \left(\Phi_1^\dag \Phi_1\right)^2 
   + \frac{1}{2} \lambda_2 \left(\Phi_2^\dag \Phi_2\right)^2  \notag\\
  +& \lambda_3 \left(\Phi_1^\dag \Phi_1\right) \left(\Phi_2^\dag \Phi_2\right)
 + \lambda_4 \left| \Phi_1^\dag \Phi_2\right |^2 \notag\\
 +& \frac{1}{2} \lambda_5 \left[  \left( \Phi_1^\dag \Phi_2 \right)^2 + \text{h.c.} \right ] \notag \\
  +& \left\{ \left[ \lambda_6 \left(\Phi_1^\dag \Phi_1\right) + \lambda_7 \left(\Phi_2^\dag \Phi_2\right)\right]
 \Phi_1^\dag \Phi_2 + \text{h.c.}\right\}.
 \label{pot2hdm} 
 \end{align}
 
 This potential has terms multiplying the parameters $m_{12}$, $\lambda_5$, $\lambda_6$ and $\lambda_7$, which in general are
 complex and, hence, are sources of $CP$ violation. The other terms in the potential are real. It is also noted that all these parameters appearing in
 the general potential are not observable, since they can be modified by a change of basis.
 
 After spontaneous EWSB, five physical Higgs
 particles are left in the spectrum: one charged Higgs pair, $H^\pm$, one CP-odd scalar, $A$, and two CP-even states, $h$ and $H$ -- where by convention $m_H>m_h$.
% $H$ (heaviest) and $h$ (lightest) given as:
% \begin{align}
% &H^\pm = \sin\beta \,\phi_1^\pm + \cos\beta \,\phi_2^\pm, \\
% &A = \sin\beta \,\text{Im} \,\phi_1^0 + \cos\beta \,\text{Im}\, \phi_2^0, \\
%&H = \cos\alpha \left(\text{Re}(\phi_1^0)-v_1\right) + \sin\alpha \left(\text{Re}(\phi_2^0)-v_2\right), \\
%&h = -\sin\alpha \left(\text{Re}(\phi_1^0)-v_1\right) + \cos\alpha \left(\text{Re}(\phi_2^0)-v_2\right).
% \end{align}
 Here $\phi_i^+$ and $\phi_i^0$ denote the $T_3=1/2$ and $T_3=-1/2$ components of the $i^{th}$ doublet for
 $i = 1, 2$. The angle $\alpha$ diagonalises the CP-even Higgs squared-mass matrix and 
 $\beta$ diagonalises both the CP-odd and charged Higgs sectors, which leads to $\tan\beta = v_2/v_1$. Note here that
 $\langle \phi_i^0 \rangle= v_i$ for $i = 1, 2$ and $v_1^2 + v_2^2 = v^2 \approx (246\,\text{GeV})^2$, where $v$ is the physical vacuum expectation value (\textit{vev}). Further choices of symmetries and couplings to quarks 
 and leptons etc. can be made, which lead to different types of models. Models which lead to natural flavour conservation are called Type-I, Type-II, 
 Lepton-specific or Flipped 2HDMs, as detailed in Ref.~\cite{Branco:2011iw}. In our studies we used a Type-II
 2HDM, upon which we added our additional scalars. 

In Ref.~\cite{cms-16-007}, a study has been carried out considering two benchmark scenarios of a 2HDM and minimal 
supersymmetric model, whereby exclusion contours are given on the model parameters using CMS Run 1 data. 
By fixing the lighter Higgs mass, 
$m_h = 125.09$~GeV, $m_A = m_H + 100$~GeV, $m_{H^\pm} = m_H + 100$~GeV, $m_H$ and $\tan\beta$ is scanned. 
The Type-I (II) 2HDM parameter space is generally constrained such that $\cos\left(\beta - \alpha \right) \lesssim 0.5 (0.2)$, 
$m_H \lesssim 380 (\approx 380)$ and 
$\tan\beta \lesssim 2$ (all). These constraints have been obtained by considering the decay channels
$A/H/h \to \tau\tau$~\cite{Chatrchyan:2014vua}, $H\to WW/ZZ$~\cite{Khachatryan:2015cwa}, $A\to ZH(llbb)$ and $A\to ZH(ll\tau\tau)$~\cite{CMS:2015mba}. 

Any extended theory beyond the SM must preserve and respect the known symmetries and constraints from theory
as well as observations from experiments. Accordingly, the following constraints apply to a 2HDM.
\begin{itemize}
\item [(a)] Vacuum stability: the Higgs potential must be bounded from below and therefore the following conditions
for $\lambda_m$ must be satisfied:
%$\lambda_1 > 0, \, \lambda_2 > 0, \, \lambda_3 > -\sqrt{\lambda_1 \lambda_2}, \notag\\
%\lambda_3 + \lambda_4 - \left| \lambda_5 \right| > -\sqrt{\lambda_1 \lambda_2}.$
\begin{align}
&\lambda_1 > 0, \, \lambda_2 > 0, \, \lambda_3 > -\sqrt{\lambda_1 \lambda_2}, 
\lambda_3 + \lambda_4 - \left| \lambda_5 \right| \notag \\ &> -\sqrt{\lambda_1 \lambda_2}. \label{vacstab}
\end{align}

\item [(b)] Perturbativity: we need the bare quartic couplings in the Higgs potential to satisfy perturbativity as
$\left|  \lambda_m\right| < 4\pi$ for $m = 1,2,..,7$. In addition, the magnitudes of quartic couplings among physical
scalars $\lambda_{\phi_i \phi_j \phi_k \phi_l}$ should also be smaller than $4\pi$, where $\phi_i = h, H, A, H^\pm$.
\item [(c)] Oblique parameters: the electroweak precision observables such as the $S$, $T$ and $U$ parameters obtain contributions 
from extra scalars in the 2HDM in loop calculations, and therefore receive contributions from $\Delta S$, $\Delta T$ and $\Delta U$.
\item [(d)] In addition, there are also some experimental constrains such as LEP bounds, flavour-changing neutral current (FCNC)
constraints, Higgs data from the LHC etc. that can restrict the model parameters.
\end{itemize} 

Recent studies on the 2HDM with its phenomenology and constraints can be found in Refs.~\cite{Coleppa:2014cca,Coleppa:2013dya,Coleppa:2013xfa}.
In general, all multi-Higgs-doublet models including 2HDMs contain the possibility of severely constrained tree level
FCNCs. To avoid these potentially dangerous interactions one can impose several
discrete symmetries in many possible ways. One such discrete symmetry to avoid FCNCs is $\mathbb{Z}_2$, which demands
invariance of the general scalar potential under the transformations $\Phi_1 \to -\Phi_1$ and $\Phi_2 \to \Phi_2$. However, this discrete
$\mathbb{Z}_2$ symmetry could be (a) {\it exact} if $m_{12}$, $\lambda_6$ and $\lambda_7$ vanish, and thus the scalar potential
will be $CP$ conserving, (b) broken {\it softly} if it is violated in the quadratic terms only, i.e., in the limit where $\lambda_6$,
$\lambda_7$ vanish, but $m_{12}$ remains non-zero and (c) {\it hard} breaking, if it is broken by the quadratic terms too, where
the parameters $m_{12}$, $\lambda_6$ and $\lambda_7$ are all non-vanishing.   

In a Type-II 2HDM the discrete $\mathbb{Z}_2$ symmetry applies for $\Phi_1 \to -\Phi_1$ and $\psi^a_R \to -\psi^a_R$, where 
$\psi^a_R$ are the charged leptons or down type quarks, and $a$ represents the generation index. However, in our studies 
the terms associated with $\lambda_6$ and $\lambda_7$ are neglected and $m_{12}$ is taken as real. 
The quadratic couplings in terms of the physical masses of the $CP$-even scalars ($m_h$, $m_H$), the $CP$-odd scalar ($m_A$)
and charged scalars ($m_{H^\pm}$) can be expressed as:
\begin{align}
&\lambda_1 = \frac{1}{v^2 \cos^2\beta} \left( m_H^2 \cos^2\alpha + v^2 m_h^2 \sin^2\alpha  - m_{12}^2 \frac{\sin\beta}{\cos\beta}\right),\\
&\lambda_2 = \frac{1}{v^2 \sin^2\beta} \left( m_H^2 \sin^2\alpha + v^2 m_h^2 \cos^2\alpha  - m_{12}^2 \frac{\cos\beta}{\sin\beta}\right),\\
&\lambda_3 = \frac{2 m_{H^+}^2}{v^2} + \frac{\sin (2\alpha)}{v^2 \sin (2\beta)}\left( m_H^2 - m_h^2 \right) - \frac{m_{12}^2}{v^2\sin\beta\cos\beta},\\
&\lambda_4 = \frac{1}{v^2} \left( m_A^2 - 2 m_{H^+}^2 \right) + \frac{m_{12}^2}{v^2\sin\beta\cos\beta},\\
&\lambda_5 = \frac{m_{12}^2}{v^2\sin\beta\cos\beta}  - \frac{m_A^2}{v^2}.
\end{align}
 
In \ref{appenprod} and~\ref{intLag2hdm} we provided the analytical expressions for production cross sections of $H$ 
and $A$, and the interaction Lagrangians in a Type-II 2HDM, respectively.

\subsection{Adding a scalar $\chi$}
\label{addx}
In order to accommodate some features in the Run 1 ATLAS and CMS results $viz.$ (a) the measurement of the differential
Higgs boson $p_\text{T}$, (b) di-Higgs resonance searches, (c) top associated Higgs production and (d) $VV$ resonance searches
(where $V = Z, W^\pm$), in Ref.~\cite{vonBuddenbrock:2015ema} it was assumed that at least one Higgs boson is 
produced due to the decay of a heavy scalar $H$ in association with a DM candidate $\chi$. However, it was explained 
in an effective theory which is briefly discussed in the next section. In this work, we consider the accommodation of $H$ in $\chi$ in a complete theory. The addition of $\chi$ as a real scalar in the 2HDM model requires additional terms in the potential 
defined in Eq.~\ref{pot2hdm}. One can consider $\chi$ as a gauge-singlet scalar and a stable DM candidate if its mixing with 
the doublets $\Phi_1$ and $\Phi_2$ can be prevented by the introduction of some discrete symmetry. One such symmetry is 
a $\mathbb{Z}_2$ under which $\chi$ is odd and all other fields are even. This also ensures the stability of $\chi$.
Thus, the most general potential consistent with the gauge and $\mathbb{Z}_2$ symmetries can be written as:
\begin{align}
{\cal V}\left(\Phi_1,\Phi_2,\chi\right) =&\,  {\cal V}\left(\Phi_1,\Phi_2\right) + \frac{1}{2}m^2_{\chi}\chi^2
 + \frac{\lambda_{\chi_{1}}}{2}\Phi_1^\dag \Phi_1 \chi^2 \notag \\
 & + \frac{\lambda_{\chi_{2}}}{2}\Phi_2^\dag \Phi_2 \chi^2  
  +\frac{\lambda_{\chi_{3}}}{4}(\Phi_1^\dag \Phi_2 + \text{h.c}) \chi^2 \notag \\
 & + \frac{\lambda_{\chi_{4}}}{8} \chi^4. \label{vchi}
\end{align}

Here we shall consider the {\it{hard}} breaking of this $\mathbb{Z}_2$ symmetry, with $\lambda_{\chi_{3}}$ being real. In the case of a {\it{soft}} breaking of the 
symmetry, the term $\lambda_{\chi_{3}}$ and corresponding terms in ${\cal V}\left(\Phi_1,\Phi_2\right)$ 
with $\lambda_6$ and $\lambda_7$ will disappear. Despite the fact that any additional scalar to the 2HDM potential
may acquire a {\it vev}, we explicitly consider the case where the additional field $\chi$ does
not acquire a {\it vev}.
Hence, in terms of the mass eigenstates, the complete interaction terms with $h$, $H$, $A$ and $H^\pm$ will be:
\begin{align}
{\cal L}_{\chi} =&\,  -\frac{1}{2} m_\chi^2 \chi^2
- \frac{1}{2} v \lambda_{h \chi \chi} h \chi^2 - \frac{1}{2} v \lambda_{H \chi \chi} H \chi^2
- \lambda_{hh\chi\chi} hh\chi^2 \notag \\ 
&- \lambda_{HH\chi\chi} HH\chi^2  - \lambda_{hH\chi\chi} hH\chi^2
- \lambda_{AA\chi\chi} AA\chi^2 \notag \\
& - \lambda_{H^+H^-\chi\chi} H^+ H^-\chi^2,
\end{align}
where the couplings are given as:
\begin{align}
\lambda_{h \chi \chi} =&\, \lambda_{\chi_{1}} \cos\beta \sin\alpha - \lambda_{\chi_{2}} \sin\beta \cos\alpha \notag \\
& - \frac{1}{2}\lambda_{\chi_{3}}\cos(\beta + \alpha), \\
\lambda_{H \chi \chi} =& -\lambda_{\chi_{1}} \cos\beta \cos\alpha - \lambda_{\chi_{2}} \sin\beta \sin\alpha \notag \\
& - \frac{1}{2}\lambda_{\chi_{3}}\sin(\beta + \alpha), \\
\lambda_{hh\chi \chi} =& \frac{1}{4}(\lambda_{\chi_1} \sin^2\alpha + \lambda_{\chi_2} \cos^2\alpha 
 - \lambda_{\chi_3} \sin\alpha \cos\alpha ), \\
\lambda_{HH\chi \chi} =& \frac{1}{4}(\lambda_{\chi_1} \cos^2\alpha + \lambda_{\chi_2} \sin^2\alpha 
 + \lambda_{\chi_3} \sin\alpha \cos\alpha ), \\
\lambda_{hH\chi\chi} =&\, \frac{1}{4}(-\lambda_{\chi_1} \cos\alpha \sin\alpha + \lambda_{\chi_2} \cos\alpha \sin\alpha \notag \\
& + \lambda_{\chi_3} \cos^2\alpha - \lambda_{\chi_3} \sin^2\alpha ), \\
\lambda_{AA\chi\chi} =& \frac{1}{4}(\lambda_{\chi_1} \sin^2\beta + \lambda_{\chi_2} \cos^2\beta - \lambda_{\chi_3} \sin\beta \cos\beta ),\\
\lambda_{H^+H^-\chi \chi} = &\frac{1}{4}(\lambda_{\chi_1} \sin^2\beta + \lambda_{\chi_2} \cos^2\beta - \lambda_{\chi_3} \sin\beta \cos\beta ) .
\end{align} 

It is also noted that ${\cal L}_\chi$ does not include $A$-$\chi$-$\chi$ interaction terms due to $CP$ violation
issues, but in principle the $CP$-odd scalar $A$ plays an important role in determining the DM relic density through
the creation or annihilation process $\chi\chi \leftrightarrow AA$. 

In addition to the constraints discussed for the 2HDM parameters, the perturbativity conditions also imply
$\left| \lambda_{\chi_m} \right| < 4\pi$ for $m=1,2,3$. The coupling $\lambda_{\chi_4}$ for the $\chi^4$ term should be
$0 < \lambda_{\chi_4} < 4 \pi$, where the lower limit is required for stability. Vacuum stability requires the following
necessary and sufficient conditions in addition to Eq.~\ref{vacstab}, so that the potential 
${\cal V}\left(\Phi_1,\Phi_2,\chi\right)$ must be bounded from below:
\begin{align}
&\lambda_{\chi_4} > 0, \\
& \lambda_{\chi_1} >  -\sqrt{\frac{1}{12} \lambda_{\chi_4} \lambda_1}, \\ 
&\lambda_{\chi_2} >  -\sqrt{\frac{1}{12} \lambda_{\chi_4} \lambda_2}, \\
&\lambda_{\chi_3} >  -\sqrt{\frac{1}{12} \lambda_{\chi_4} \lambda_3}. \label{vacstabchi}
\end{align}

If $\lambda_{\chi_1}, \lambda_{\chi_2}, \lambda_{\chi_3} < 0$, then the additional conditions should also satisfy:
\begin{align}
& -2\lambda_{\chi_1}\lambda_{\chi_2} + \frac{1}{6} \lambda_{\chi_4} \lambda_3 \notag \\
& >  -\sqrt{4 \left( \frac{1}{12} \lambda_{\chi_4} \lambda_1 - \lambda_{\chi_1}^2 \right) 
\left( \frac{1}{12} \lambda_{\chi_4} \lambda_2 - \lambda_{\chi_2}^2 \right) }, \\  
& -2\lambda_{\chi_1}\lambda_{\chi_2} + \frac{1}{6} \lambda_{\chi_4} \left(\lambda_3+\lambda_4 - \left| \lambda_5\right| \right) \notag \\
& >  -\sqrt{4 \left( \frac{1}{12} \lambda_{\chi_4} \lambda_1 - \lambda_{\chi_1}^2 \right) 
\left( \frac{1}{12} \lambda_{\chi_4} \lambda_2 - \lambda_{\chi_2}^2 \right) }. \label{vacstabchiadd}
\end{align}

In order to ensure a stable DM candidate $\chi$, we need to have an additional condition that the \textit{vev}, $\langle \chi \rangle$, should vanish at the global minimum of the scalar potential in Eq.~\ref{vchi}. This can be obtained
numerically such that $\langle \chi \rangle = 0$, $\langle \Phi_1 \rangle \neq 0$ and $\langle \Phi_2 \rangle \neq 0$.  
Practical studies and analyses on the model follow these constraints  with $m_\chi < m_h/2$.     
In Ref.~\cite{Drozd:2014yla} a similar study can be seen.

In this work we consider $\chi$ to be a scalar. However, while considering various features in the data, this may not be an appropriate assumption. In light of this, it is important to characterise $\chi$ in terms of other possible theories. This could shed light on the production mechanisms and decay modes for $H$ and $A$ through $gg$ and $\gamma\gamma$, since they are loop induced processes. It is possible for $\chi$ to run in these loops, and this could explain an enhancement of these rates. 
This would imply that $\chi$ is a massive coloured fermion. 

Simple possibilities for these extra fermions may be:
\begin{itemize}
	\item a single vector-like quark of charge 2/3,
	\item an isospin doublet of vector-like quarks of charges 2/3 and -1/3,
	\item an isodoublet and two singlet quarks of charges 2/3 and -1/3, or
	\item a complete vector-like generation including leptons as well as quarks.
\end{itemize}
In this respect we should 
consider all four possible characteristics of $\chi$ being a vector-like fermion (VLF). Similar studies can follow for the $W^\pm$ and
$Z$ related decay modes of $A$. 

\begin{figure*}[t]
	\centering 
	\includegraphics[height=100pt]{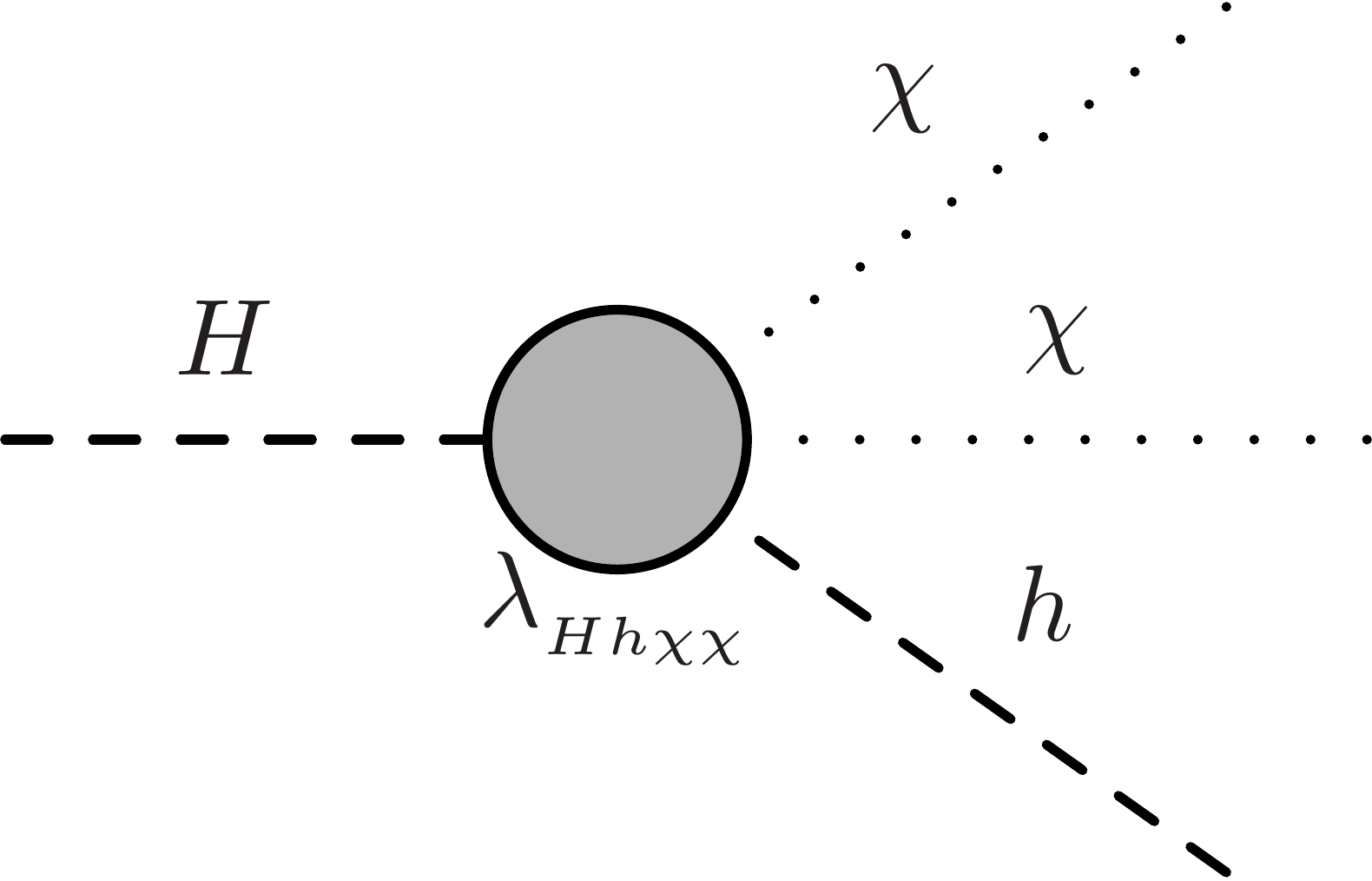}
	~~~~~~~~~~~~~~
	\includegraphics[height=103pt]{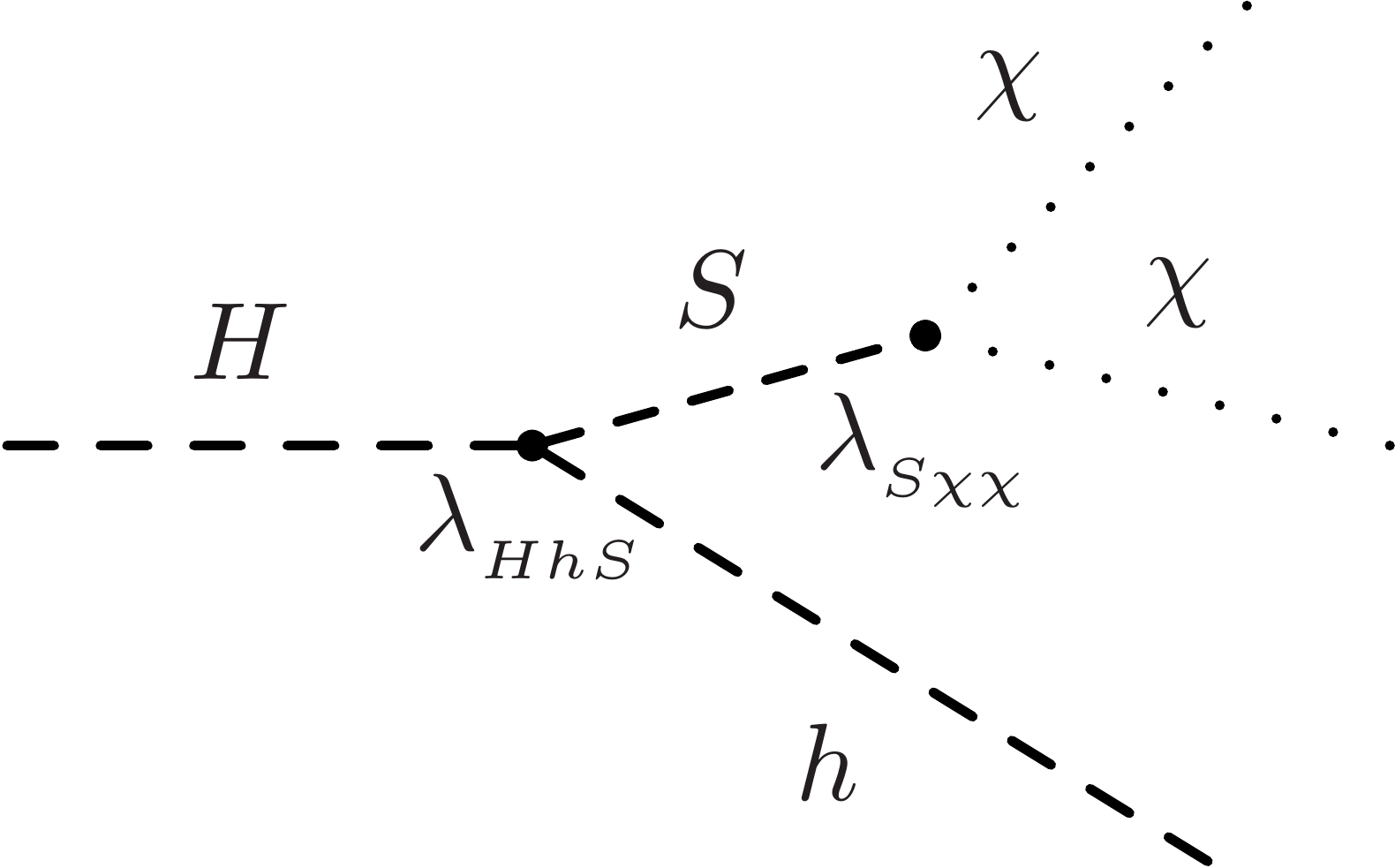} 
	\caption{\label{fig:i} Representative Feynman diagrams to study Higgs boson $p_\text{T}$ spectrum using the effective Lagrangian
		approach described by Eqs.~\ref{lag} and \ref{lagt}. On the left, through the quartic $\lambda_{hH\chi\chi}$ vertex and
		on the right due to an additional scalar $S$, as described in text. Equivalence between two procedures can explain the strength of 
		the coupling $\lambda_{Hh\chi\chi}$ under a replacement with $\lambda_{HhS}$ and $\lambda_{S\chi\chi}$.}
\end{figure*}

\subsection{Adding a Higgs-like $CP$-even scalar $S$}
\label{adds}

Previously we discussed the inclusion of a real scalar $\chi$ and accordingly
its new interactions will appear in a 2HDM. Similarly, one can 
introduce a real scalar $S$, which is chosen to be similar to the SM Higgs boson with the allowed mass range $m_S \in [m_h, m_H-m_h]$. $S$ was initially introduced as a mediator to explain the $H\to h\chi\chi$ decay mode, as shown in Fig.~\ref{fig:i}, however it can be used to probe more interesting physics. For simplicity we can impose a $\mathbb{Z}_2$ symmetry for $S \to -S$ transformations, but this can also
be relaxed for other implications in the theory. While introducing $\chi$ in the 2HDM, we only consider its couplings
with the scalars of this model i.e. $h$, $H$, $A$ and $H^\pm$. But in the case of $S$, which is SM Higgs-like, it
is allowed to couple with all of the SM particles as well as $\chi$. This is phenomenologically interesting for two reasons. Firstly, $S$ can be thought of as a portal between which SM particles can interact with DM. Secondly, the Higgs-like nature of $S$ drastically reduces the number of free parameters in the theory, since all of the BRs to SM particles (and hence coupling strengths) are fixed to what a SM Higgs boson would have, scaled down appropriately by the introduction of an invisible decay mode $S\to\chi\chi$. Since a large invisible BR is not experimentally observed for $h$, we can rather explore DM interactions with $S$.

It is clear that in the absence of such interactions, one should not expect any
interesting physics. But mixing with SM particles along with other scalars of the 2HDM has two different consequences.
Firstly, $S$ could be observed as a resonance through $p p \to S \to VV$ modes, where $V = Z, W^\pm,~\gamma$. For a Higgs-like $S$, such searches would be similar to generic Higgs boson searches at higher masses, and the signal and background modelling would therefore be the same. However, it should be noted that in this study we consider direct production of $S$ to be small, and $S$ is produced dominantly through the decay of $H$.
Secondly, it alters the coupling strengths of known interactions in the theory --
for example, in a 2HDM there follows a sum rule for the neutral scalar gauge couplings $g_{hWW}^2 + g_{HWW}^2$, which is
the same as the SM coupling squared~\cite{Chen:2013jvg}. This sum rule will be violated if there is any mixing occurring between $S$
and the doublets $\Phi_{1,2}$, which will directly alter the expected projected bounds of 2HDM couplings.
    
In light of this, we add a real\footnote{One can also introduce a complex scalar in theory, the consequence of which
alters the choice of symmetry. The $\mathbb{Z}_2$ symmetry would then be promoted to a global $U(1)$ and its spontaneous
breaking would lead to a massless pseudoscalar.} 
scalar $S$ considering the possibility of a discrete symmetry under $S \to -S$. The parameters are arranged
in such a way so that $S$ acquires a {\it vev}. Without the discrete symmetry, the most general potential for $S$ can be written as:
\begin{align}
{\cal V}\left(\Phi_1,\Phi_2,S\right) =&\,  {\cal V}\left(\Phi_1,\Phi_2\right) + \frac{1}{2}m^2_{S_0}S^2
 + \frac{\lambda_{S_{1}}}{2}\Phi_1^\dag \Phi_1 S^2 \notag \\
 & + \frac{\lambda_{S_{2}}}{2}\Phi_2^\dag \Phi_2 S^2 
 + \frac{\lambda_{S_{3}}}{4}(\Phi_1^\dag \Phi_2 + \text{h.c}) S^2 \notag \\
& + \frac{\lambda_{S_{4}}}{4!} S^4 
 + \mu_{1}{\Phi^\dagger_1}{\Phi_1}S + \mu_{2}{\Phi^\dagger_2}{\Phi_2}S \notag \\
& + \mu_{3}\left[{\Phi^\dagger_1}{\Phi_2} 
 + \rm h.c \right]S + \mu_{S}S^3.
\end{align}

Now, if we impose a $\mathbb{Z}_2$ symmetry for transformations of the form $S \to -S$ (and all other fields are even), then the terms with the coefficient $\mu_i$ ($i = 1, 2, 3, S $) will vanish in the above
general potential. If we further assume another $\mathbb{Z}_2^{'}$ symmetry for the transformations
$h\to h$, $H\to -H$ and $S\to S$, then the $\lambda_{S_3}$ term will also vanish. This also
eliminates $\lambda_6$ and $\lambda_7$ from ${\cal V}\left(\Phi_1,\Phi_2\right)$. However,
we assume a soft breaking of $\mathbb{Z}_2^{'}$, which allows $m_{12}^2\neq 0$. In the case where $S$ does not acquire a {\it vev} (similar to $\chi$). Then the 
$S$ related interactions in the potential are given by:
\begin{align}
V_S = & \frac{1}{2}m_S^2S^2 + \lambda_{hSS}v~hS^2 + \lambda_{HSS}v~HS^2 - 
\lambda_{HHSS}~H^2S^2 \nonumber\\
& - \lambda_{hHSS}~hHS^2 - \lambda_{hhSS}~h^2S^2 \notag \\
& - \lambda_{AASS}~A^2S^2 -\lambda_{H^+H^-SS}~H^+H^-S^2.
\end{align}

One can write various couplings in the potential in terms of $\lambda_{S_1}$, $\lambda_{S_2}$,
$\alpha$ and $\beta$ as follows:
\begin{align}
&m_S^2 = m_{S_0}^2 + \left(\frac{\lambda_{S_1}}{2}\cos^2\beta+\frac{\lambda_{S_2}}{2}\sin^2\beta\right)v^2 \\
&\lambda_{hSS} = -\frac{\lambda_{S_1}}{2}\sin\alpha\cos\beta+\frac{\lambda_{S_2}}{2}\cos\alpha\sin\beta \label{eq:lhss} \\
&\lambda_{HSS} = \frac{\lambda_{S_1}}{2}\cos\alpha\cos\beta+\frac{\lambda_{S_2}}{2}\sin\alpha\sin\beta \label{eq:lh2ss} \\
&\lambda_{hhSS} = \frac{\lambda_{S_1}}{4}\sin^2\alpha+\frac{\lambda_{S_2}}{4}\cos^2\alpha \\
&\lambda_{HHSS} = \frac{\lambda_{S_1}}{4}\cos^2\alpha+\frac{\lambda_{S_2}}{4}\sin^2\alpha \\
&\lambda_{hHSS} = \frac{1}{4}\left(\lambda_{S_2} - \lambda_{S_1}\right)\sin 2\alpha \\
&\lambda_{AASS} = \frac{1}{2}\lambda_{H^+H^-SS} = \frac{\lambda_{S_1}}{4}\sin^2\beta+\frac{\lambda_{S_2}}{4}\cos^2\beta. \label{eq:laass}
\end{align}

In order to generate an effective $hH\chi\chi$ type interaction from a full model with
$S$, we need to allow a coupling $hHS$. This coupling can be generated from the $hHSS$
interaction if $S$ acquires a {\it vev}. Therefore, 
the $S$ in our model will indeed acquire a {\it vev} and mix 
with $h$ and $H$.

From Ref.~\cite{Khachatryan:2016vau}, one can infer that $hS$ mixing must be small if it exists, with an upper limit on the mixing squared at about 20\%. In the limit of zero mixing between $h$ and $S$ (as well as $H$ and $S$), the expressions for various couplings are shown in Eqs.~\ref{eq:lhss}-\ref{eq:laass}. Eq.~\ref{eq:lh2ss} tells us that the $HSS$ coupling need not be small even in this limit, since $\alpha$ and $\beta$, which are the mixing angles from the doublet sector exclusively, are free parameters. If we turn on a mixing between $S$ and the doublets, Eq.~\ref{eq:lh2ss} will receive corrections through the additional mixing angle(s) introduced. However, in case of small $hS$ mixing, the correction will also be small, and the $HSS$ coupling will still remain sizeable.

Therefore, we assume that the mixing of $S$ with $h$ is small enough 
(by interplay of various parameters in the potential) that it will not spoil any experimental bounds. 
The $hHSS$ interaction can be thought of as a source of the required $hHS$ coupling if
we replace one $S$ by its {\it vev} in the $hHSS$ interaction.   

\section{An effective theory approach to explain the Higgs $p_\text{T}$ spectrum}
\label{hpteff}

To explain distortions in the Higgs boson $p_\text{T}$ spectrum, we can consider an effective Lagrangian approach with the introduction of two hypothetical real scalars, 
$H$ and $\chi$, which are beyond the SM (BSM) in terms of its particle spectrum -- as discussed in Ref.~\cite{vonBuddenbrock:2015ema}. 
This effective model can also be used to study other phenomenology associated with Higgs physics.  
The formalism considers heavy scalar boson production though gluon-gluon fusion ($gg$F), which then decays into the SM Higgs and a pair of $\chi$ particles. As before, $\chi$ is considered as a DM candidate and therefore a source of missing transverse energy (\MET). 

The required vertices for these studies are:
\begin{align}
\mathcal{L}_{H} =& -\frac{1}{4}~\beta_{g} \kappa_{_{hgg}}^{\text{SM}}~G_{\mu\nu}G^{\mu\nu}H
+\beta_{_V}\kappa_{_{hVV}}^{\text{SM}}~V_{\mu}V^{\mu}H,  \label{vh} \\
\mathcal{L}_{\text{Y}} =& -\frac{1}{\sqrt{2}}~\Big[y_{_{ttH}}\bar{t} t H + y_{_{bbH}} \bar{b} b H\Big],\\ 
\mathcal{L}_{\text{T}} =& -\frac{1}{2}~v\Big[\lambda_{_{Hhh}}Hhh + \lambda_{_{h\chi\chi}}h\chi\chi + \lambda_{_{H\chi\chi}}H\chi\chi\Big], \\
\mathcal{L}_{\text{Q}} =& -\frac{1}{2}\lambda_{_{Hh\chi\chi}}Hh\chi\chi - \frac{1}{4} \lambda_{_{HHhh}}HHhh 
-\frac{1}{4}\lambda_{_{hh\chi\chi}}hh \chi\chi \notag \\
& - \frac{1}{4} \lambda_{_{HH\chi\chi}}HH\chi\chi, \label{vq}
\end{align} 
where $\beta_g = y_{ttH}/y_{tth}$ is the scale factor with respect to the SM Yukawa top coupling for $H$, and it is therefore
used to tune the effective $gg$F coupling. A similar factor $\beta_V$ is used for $VVH$ couplings. 
The complete set of these new interactions are added to the SM Lagrangian, ${\cal{L}_{\text{SM}}}$, and thus the final Lagrangian is 
$\mathcal{L} = \mathcal{L}_{\text{SM}} + \mathcal{L}_{\text{BSM}}$,
where $\mathcal{L}_{\text{BSM}}$ contains the terms beyond the SM interactions which is given by 
\begin{align}
\mathcal{L}_{\text{BSM}} =&\, \frac{1}{2} \partial_\mu \chi \partial^\mu \chi - \frac{1}{2} m_\chi^2 \chi \chi
 + \frac{1}{2} \partial_\mu H \partial^\mu H \notag \\
 & - \frac{1}{2} m_H^2 H H 
 + \mathcal{L}_{H} + \mathcal{L}_\text{Y} + \mathcal{L}_\text{T} + \mathcal{L}_\text{Q}.\label{lag}
\end{align}

Here, we should note that $\chi$ only interacts with the SM Higgs and the postulated heavy scalar $H$ -- not with the SM fermions and gauge bosons. We also require that $\chi$ is stable by imposing
the appropriate symmetry conditions which we described previously in Sect.~\ref{addx}.
Since we assume $\chi$ to be a DM candidate,      
there are non-negligible constraints on the associated parameters of the vertices that come from the relic density of DM and the DM-nuclei 
inelastic scattering cross sections. In addition to this, constraints arise from limits on the invisible BR of the SM Higgs boson. These leave a narrow choice of the mass of the DM candidate, $m_\chi \sim m_h/2$, as well as the parameter $\lambda_{h\chi\chi} \sim [0.0006-0.006]$. We also assume that $m_H$ would lie in the range, 
$2 m_h < m_H < 2 m_t$ to forbid the $H\to t t$ decay, as well as keep the $H\to h\chi\chi$ decay on-shell. 

\begin{figure}
	\includegraphics[width=0.45\textwidth]{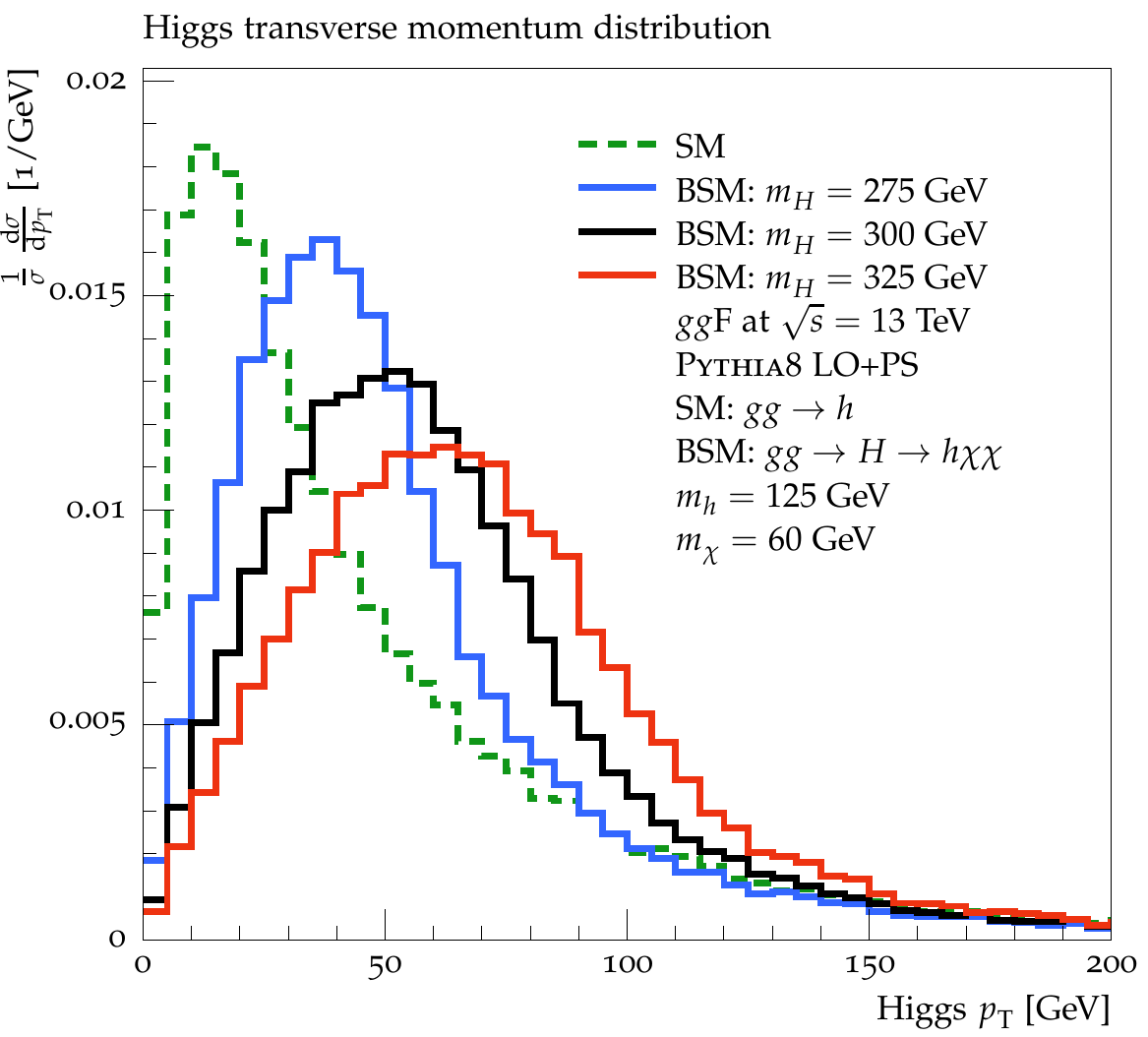}
	\caption{The impact of the effective decay process $H\to h\chi\chi$ on the Higgs $p_\text{T}$ spectrum. Under the BSM hypothesis of $gg\to H\to h\chi\chi$ (solid lines), the spectrum is distorted with respect to the SM prediction (dashed line). These distributions were made using 50~000 events generated at leading order (LO) and showered in \textsc{Pythia8}~\cite{Sjostrand:2014zea}. Three mass points of $H$ are chosen for demonstration, and $m_\chi=60~\text{GeV}\sim m_h/2$.}
	\label{fig:higgspt}
\end{figure}

\begin{figure}
	\includegraphics[width=0.45\textwidth]{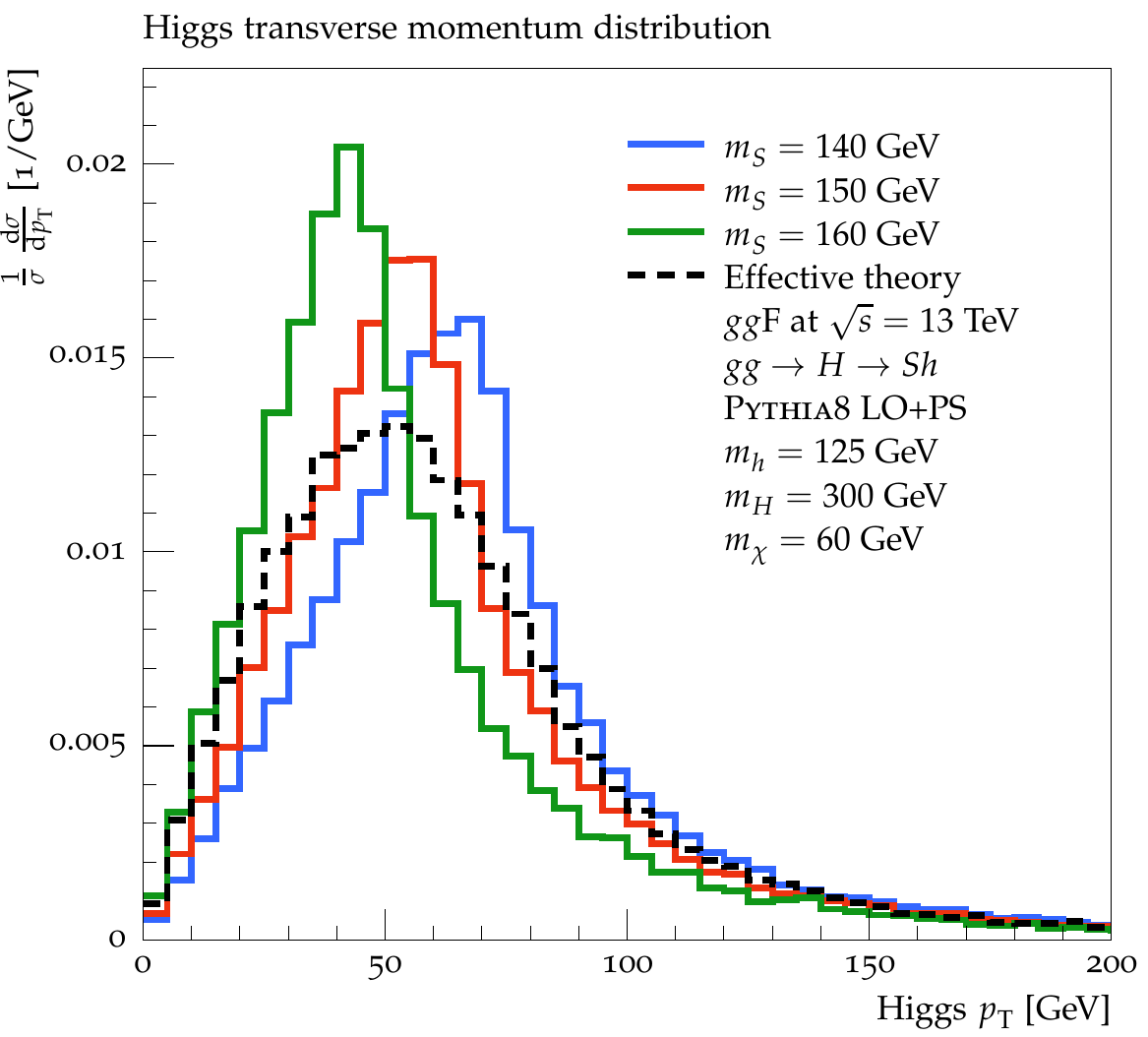}
	\caption{A comparative view of the Higgs $p_\text{T}$ spectrum as described by the $S$-mediated interaction (solid lines) and the effective $H\to h\chi\chi$ interaction (dashed line). In this case, $m_H$ is fixed to 300~GeV and $m_S$ is varied. The generator setup is similar to that in Fig.~\ref{fig:higgspt}.}
	\label{fig:model_compare}
\end{figure}

In this study, if we consider the process $pp \to H \to h\chi\chi$, then a distortion could be predicted in the intermediate range of the
Higgs $p_\text{T}$ spectrum. This comes from the recoil of $h$ against a pair of invisible $\chi$ particles, and the effect on the Higgs $p_\text{T}$ spectrum can be seen in Fig.~\ref{fig:higgspt}. On introducing the $S$ to mediate the effective interaction, the kinematics for the effective theory will be similar to the full theory with a large width $S$ at $m_S=m_H/2$ (in the limit $m_\chi\to m_h/2$). The Higgs $p_\text{T}$ spectrum arising from the $S$-mediated interaction can be seen in Fig.~\ref{fig:model_compare}; three mass points have been chosen to demonstrate the effect of $m_S$ on the spectrum.
In order to chose appropriate values of associated couplings one must consider the constraints from all potential experimental signatures which the model predicts i.e. di-Higgs and di-boson production through the resonance $H$, and top associated $H$ production (in comparison to top associated $h$ production) etc.   

In an effective field theory approach, we do not consider the actual origin of the $Hh\chi\chi$ coupling. One can assume 
that this effective interaction is mediated by the scalar particle $S$ which will then decay in the mode $S\to\chi\chi$. This inclusion of $S$ can 
open up various new possibilities in terms of search channels and phenomenology. In addition to the above studies, if we look over the di-Higgs production modes 
in different decay channels (such as$\gamma\gamma b \bar b$ or $b\bar b b \bar b$ with jets etc.), then the vertices defined above
(in Eqs.~\ref{vh} to \ref{vq}) will be modified appropriately with $S$ as an intermediate scalar and not as a DM candidate.\footnote{$S$ is a scalar particle with various decay modes, therefore having all possible branchings to other particles. As a result, the symmetry requirements for a gauge invariant set of vertices in the 
Lagrangian is different.}
With the mass range $m_h \lesssim m_S  \lesssim m_H - m_h$ and $m_S > 2m_\chi$, new possibilities for the processes in these studies include $pp \to H\to hS$ as well as $pp \to H \to hh$, considering the available
spectrum of $m_S$ and the associated coupling parameters.
There is a possibility to introduce a $HSS$ vertex in the study, which participates further in a $H \to S S$ 
decay channel (similar to $H \to h h$).
An important feature to keep in mind is that all decay modes of $S$ (i.e. $S$ into jets, vector bosons, leptons, DM etc.) are possible. 

Following the effective theory approach, and after EWSB, the Lagrangian for singlet real scalar 
$S$ can be written as:
\begin{equation}
{\cal L}_{S} = {\cal L}_\text{K} + {\cal L}_{SVV^\prime}
 + {\cal L}_{Sf\bar f} + {\cal L}_{hHS} + {\cal L}_{S\chi},
\end{equation}
where
\begin{equation}
{\cal L}_\text{K} =\, \frac{1}{2} \partial_\mu S \partial^\mu S - \frac{1}{2} m_S^2 S S, \label{lsk}
\end{equation}
\begin{align}
{\cal L}_{SVV^\prime} =&\, \frac{1}{4} \kappa_{_{Sgg}} \frac{\alpha_{s}}{12 \pi v} S G^{a\mu\nu}G_{\mu\nu}^a
+ \frac{1}{4} \kappa_{_{S\gamma\gamma}} \frac{\alpha}{\pi v} S F^{\mu\nu}F_{\mu\nu} \notag \\
& + \frac{1}{4} \kappa_{_{SZZ}} \frac{\alpha}{\pi v} S Z^{\mu\nu}Z_{\mu\nu} 
  + \frac{1}{4} \kappa_{_{SZ\gamma}} \frac{\alpha}{\pi v} S Z^{\mu\nu}F_{\mu\nu} \notag \\
& + \frac{1}{4} \kappa_{_{SWW}} \frac{2 \alpha}{\pi s_w^2 v} S W^{+\mu\nu}W^{-}_{\mu\nu}, \label{lsvv}
\end{align}
\begin{equation}
{\cal L}_{Sf\bar f} =\, - \sum_f \kappa_{_{Sf}} \frac{m_f}{v} S \bar f f, \label{lsf}
\end{equation}
\begin{align}
{\cal L}_{HhS} =&\, -\frac{1}{2}~v\Big[\lambda_{_{hhS}} hhS + \lambda_{_{hSS}} hSS + 
\lambda_{_{HHS}} HHS \notag \\ & + \lambda_{_{HSS}} HSS + \lambda_{_{HhS}} HhS\Big], \label{lsh}
\end{align}
\begin{equation}
{\cal L}_{S\chi} =\, -\frac{1}{2}~v~\lambda_{_{S\chi\chi}} S\chi\chi -\frac{1}{2} \lambda_{_{SS\chi\chi}} SS\chi\chi. \label{lsc} 
\end{equation}

Here $V, V^\prime \equiv g, \gamma, Z \,\text{or}\, W^\pm$ and $W^\pm_{\mu\nu} = D_\mu W^\pm_\nu - D_\nu W^\pm_\mu $,
$D_\mu W^\pm_\nu = \left[ \partial_\mu \pm i e A_\mu \right] W^\pm_\nu$. Other possible self interaction terms for $S$
are neglected here since they are not of any phenomenological interest for our studies. Hence the total effective Lagrangian is:
\begin{equation}
\cal{L}_{\text{tot}} = \cal{L}_\text{SM} + \cal{L}_\text{S}. \label{lagt}
\end{equation} 

\begin{table*}[t]
	\renewcommand{\arraystretch}{1.15}
	\centering
	\begin{tabular}{c|c|l}
		\hline
		\textbf{S. No.}                 & \textbf{Scalars}       & \textbf{Decay modes} \\ \hline
		\textsc{D.1} & $h$ & $b\bar b$, $\tau^+ \tau^-$, $\mu^+\mu^-$, $s\bar s$, $c\bar c$, $gg$, $\gamma\gamma$, $Z\gamma$,
		$W^+W^-$, $ZZ$     \\
		\textsc{D.2} & $H$ & \textsc{D.1}, $hh$, $SS$, $Sh$ \\ %, $h\chi\chi$      \\
		\textsc{D.3} & $A$ & \textsc{D.1}, $t\bar t$, $Zh$, $ZH$, $ZS$, $W^\pm H^\mp$      \\
		\textsc{D.4} & $H^\pm$ & $W^\pm h$, $W^\pm H$, $W^\pm S$      \\
		\textsc{D.5} & $S$ & \textsc{D.1}, $\chi\chi$      \\
		\hline
	\end{tabular}
	\caption{\label{tab:i} The list of possible decay modes of the 2HDM scalars and $S$
		based on the explicit mass choices as described in the text. Note that we are not interested in $h\to \chi\chi$
		decay; instead we prefer $S\to \chi\chi$ decay mode.}
\end{table*}
 
It is interesting to note that the choice of narrow mass range for $S$, $m_S\in [m_h, m_H-m_h]$ 
provides an opportunity to see various phenomenological aspects of the model in contrast to $h$. A few examples include the $S\to \chi\chi$
mode that predicts \MET in Higgs-like events, monojet searches through $Sj$, or di-jet events in association with \MET through $S+\text{jets}$ decays. The mass range for $S$ may help to understand
rates for a Higgs-like scalar in different possible production or decay modes too. 
An important search (after the SM Higgs discovery) at the LHC could be for a scalar candidate $S$ 
through resonance production in either of the di-boson decay channels, $S \to VV,$ and $S\to \gamma\gamma$.

If we perform more investigation on the effective terms considered in above set of Lagrangians (most notably ${\cal L}_{_{HhS}}$),
then the terms $hhS$, $hSS$ and $HHS$ are less relevant for the phenomenology due to the choice of a narrow mass window
of $S$. However, the two terms with $HSS$ and $HhS$ are important. The origin for the consideration of the intermediate real 
scalar $S$ demands that these two terms can explain the large BR of $H \to h\chi\chi$.   
In one sense, there is an equivalence of the couplings $\lambda_{_{Hh\chi\chi}}$ with the cascade of $\lambda_{_{HhS}}$ and $\lambda_{_{S\chi\chi}}$,
so that the 3 body decay can be equated to a series of 2 body decays, as shown in Fig.~\ref{fig:i}.
On the other hand, in order to minimise the number of free parameters in the theory, we consider a ratio of 
couplings $r = {\left| \lambda_{_{HSS}} \right|}/{\left| \lambda_{_{HhS}} \right|}$. This ratio\footnote{We make sure 
the ratio $r$ is positive definite so that there won't be any negative interference due to the choice of negative values
of couplings $\lambda_{_{HSS}}$ or $\lambda_{_{HhS}}$.} could be fixed in the limits of theoretically
allowed values, and then either one of the couplings $\lambda_{_{HSS}}$ or $\lambda_{_{HhS}}$ can be varied to control the
rates of the processes which are studied.

\section{Phenomenology}
\label{pheno}

The phenomenology discussed in the previous section (i.e. with $H$, $S$ and $\chi$ in an effective theory) can also be studied 
in a model-dependent way. In Sect.~\ref{theory} we discussed the particle spectrum of a 2HDM with two real singlet scalars and their
interactions in Type-II 2HDM scenarios, while also considering a specific $\mathbb{Z}_2$ symmetry. Here we discuss various phenomenology associated with this particle spectrum applicable to collider
signatures (in particular at the LHC). Given the mass range of each new scalar, their appropriate dominant decay modes have been listed in Table~\ref{tab:i} as a reference for the discussion on the experimental signatures. An explicit list of experimental search channels is presented in Table~\ref{tbl:los}.

\subsection{Heavy scalar $H$}
\label{hsh}
In Sect.~\ref{hpteff},  a heavy scalar $H$ was introduced in an effective theory, with the primary goal explaining a distortion in the
$p_\text{T}$ spectrum of the Higgs boson. Considering the analyses performed with the effective theory approach, 
we can now think of $H$ as the heavier CP even component
of a 2HDM.\footnote{It should be noted that in the effective Lagrangian discussed in Sect.~\ref{hpteff}, the scalar $H$ need not be a 2HDM heavy scalar.} Furthermore, our motive
should then be to fit parameters such as $\tan\beta$, $\alpha$ and the masses of $A$ and $H^\pm$ in this specific
model. However, the question arises as to whether we should think of a generalised 2HDM or any particular
type of this model, as described in detail in Ref.~\cite{Branco:2011iw}. On the other hand, we also need to consider
experimental data from searches, which will affect the possible processes taken into consideration using
this model.

Note that in this study, we explicitly choose that the lighter CP even component of a 2HDM is the experimentally observed scalar (i.e. $m_h = 125$~GeV). With this fixed, we choose the $H$ mass to be in the range $2 m_h < m_H < 2 m_t$ for reasons which were explained in Sect.~\ref{hpteff}.  

In the simplest case, the cross section of $gg\to H$ production (i.e. the dominant production mode) would be the same as a heavy Higgs boson -- between 5 and 10~pb at $\sqrt{s}=13$~TeV~\cite{Heinemeyer:2013tqa}. However, this number could be altered if one considers a rescaling of the Yukawa coupling or the possibility of extra coloured particles running in the loop (as alluded to above). In Ref.~\cite{vonBuddenbrock:2015ema}, the number $\beta_g$ -- which was assumed as a rescaling of the Yukawa coupling -- was estimated to be around 1.5. This implies that the $gg\to H$ production cross section could be enhanced by as much as a factor of 2.

\subsection{CP-odd scalar $A$}
\label{hsa}

Typically, experimental resonance searches hope to see excesses around a particular mass range (with the appropriate decay 
width approximation) 
in the invariant mass spectra of di-jet or di-boson final states. These spectra provide hints for new BSM particles to be discovered. 
The masses of these resonances $m_\Phi$ (where for a 2HDM $\Phi = H, A , H^\pm$) 
might be of the order of $2 m_h < m_\Phi < 2 m_t$ 
(which we considered in our previous studies for $m_H$) or 
beyond this order -- perhaps $2 m_t \ll m_\Phi < \mathcal{O}$(1 TeV) or even $m_\Phi \gg \mathcal{O}$(1 TeV).

In terms of phenomenological aspects for a 2HDM CP-odd scalar $A$, the following salient features could be observed:
\begin{itemize}
 \item [(1)] In 2HDMs masses of $A$ and $H^\pm$ are correlated. 
 So if we wish to have a 2HDM with a particular mass $m_A$, its compatibility with $m_{H^\pm}$ should also be
 considered. With a known value of $m_H$ ($2 m_h < m_H < 2 m_t$) and $m_h = 125$ GeV, one should tune the parameters 
 $\alpha$ and $\beta$ accordingly.
 \item [(2)] In the case of $gg$F production for $A$ (through the $ggA$ vertex), there will be a need for a scaling factor 
 $\beta_g^A$ (in a similar way to the treatment of $H$ production, which scales with $\beta_g$). Considering the decay modes of $A$, $A \to \gamma\gamma$ 
 in particular needs another scaling factor $\beta_\gamma^A$. In this respect, one needs to control the 
 $H\to \gamma\gamma$ 
 decay rates via another parameter $\beta_\gamma$, since the form factors appearing in the calculation of $gg \to H, A$ and
 $H, A \to \gamma\gamma$ have a different structure. They are also dependent on the masses of the particles under consideration
 (this is described in Refs.~\cite{gunion, Kumar:2016vut}). One should also study other possible decay modes of $A$ which include pairs of $W^\pm$ or 
 $Z$ bosons in the final state. These decays are possible only at loop level in 2HDMs, since $AW^+W^-$ and $AZZ$ couplings are absent as a result of CP conservation issues.
 \item[(3)] Depending on parameter choices, this model can predict an arbitrarily large amount of $Z+$jets$+$\MET events. It is important to think of the contribution of the
 decay mode of $A \to Z H$, where $H\to h\chi\chi$. This requires that
 $m_A > m_Z + m_H$.
\item [(4)] With respect to point (3), we can also consider different processes with multi-lepton final states  
through same-sign and opposite-sign lepton selection, in association with jets. This phenomenological interest arises from the inclusion of the
charged bosons, $H^\pm$.
 \item[(5)] Since the SM Yukawa couplings for top quarks, $y_{tth}$, are well known, one will need to adjust the parameters $\alpha$ and
 $\beta$ in such a way so that $y_{ttA}$ and $y_{ttH}$ must follow the appropriate branchings for $A\to t \bar t$
 and $H \to t \bar t$. It should be noted here that since $y_{tth}$ is close to unity (due to large top-quark mass),
 it can also add insight into new physics scales.     
\end{itemize}

\subsection{Charged scalars $H^\pm$}
\label{hshpm}

In the 2HDM particle spectrum, we also have the possibility of charged bosons, $H^\pm$, which can be produced at the LHC. Searches for these particles most often consider production cross sections and BRs in different decay
channels. 
The prominent decay modes of $H^{\pm}$ are $H^\pm \to tb$ and $H^\pm\to W^\pm h$ when $m_{H^\pm} > m_t$. 
Since we consider $2 m_h < m_H <  2 m_t$, the decay mode of $H^\pm \to W^\pm H$ could then be a prominent 
channel too in the case of $m_{H^\pm} \gg m_H$. 

The phenomenological features of $H^\pm$ are a subject of some detail, since one could consider either $m_{H^\pm} < m_t$ 
or $m_{H^\pm} > m_t$. Due to this fact, the decay modes for our studies are largely dependent on $m_{H^\pm}$, following the
appropriate mixing parameters $\alpha$ and $\beta$. We explicitly consider the case in which $m_{H^\pm} > m_t$.
The production of $H^\pm$ at the LHC would then follow two production mechanisms which can have sizeable production cross sections. These are:
\begin{itemize}
	\item $2 \to 2$, $p p \to gb(g\bar b) \to tH^- (\bar tH^+)$, and
	\item $2 \to 3$, $p p \to gg/qq^\prime \to t H^- \bar b + \bar t H^+ b$.
\end{itemize}
Additionally, $H^\pm$
production at hadron colliders can be studied through Drell-Yan like processes for pair production (i.e. $qq \to H^+H^-$). Similarly, the associated
 production with $W$ bosons ($qq \to H^\pm W^\pm$), and pair production through $gg$F
 can also be studied. 

The prominent decay modes for $H^\pm$ are $H^\pm \to tb$, $H^\pm \to \tau \nu$ and $H^\pm \to W^\pm h$. With the
allowed vertices in the 2HDM, one could think of channels where $H^\pm$ couples with $H$ (and thereafter $H \to h \chi\chi$).
This allows us to to study a final state in terms of $\chi$. Therefore, the decay mode $H^\pm \to W^\pm H$ can be highlighted in these studies as a prominent channel.
The phenomenology of $H^\pm$ also depends on whether (i) $m_h < m_H < m_A$ or (ii) $m_h < m_A < m_H$, 
since $m_{H^{\pm}}$ could be considered as heavy as $m_A$. 

\begin{table*}[t]
	\renewcommand{\arraystretch}{1.15}
	\centering
	\begin{tabular}{c|l|l}
		\hline
		\textbf{Scalar} & \textbf{Production mode} & \textbf{Search channels} \\
		\hline
		\multirow{9}{10pt}{$H$} & $gg\to H, Hjj$ ($gg$F and VBF) & Direct SM decays as in Table~\ref{tab:i} \\
		& & $\to SS/Sh\to 4W\to 4\ell$ + \MET \\
		& & $\to hh\to \gamma\gamma b\bar{b},~b\bar{b}\tau\tau,~4b,~\gamma\gamma WW$ etc. \\
		& & $\to Sh$ where $S\to\chi\chi\implies \gamma\gamma,~b\bar{b},~4\ell$ + \MET \\
		\cline{2-3}
		& $p p \to Z (W^\pm) H~(H\to SS/Sh)$ & $\to$ $6(5) l$ + \MET \\
		& & $\to 4(3) l + 2j$ + \MET \\
		& & $\to 2(1) l + 4j$ + \MET \\
		\cline{2-3}
		& $p p \to t \bar t H, (t+\bar t)H~(H \to S S / Sh)$ & $\to2 W + 2 Z$ + \MET and $b$-jets \\
		& & $\to6W\to3~\text{same sign leptons}$ + jets and \MET \\
		\hline
		\multirow{4}{10pt}{$H^\pm$} & $p p \to t H^\pm~(H^\pm\to W^\pm H)$ & $\to6W\to3~\text{same sign leptons}$ + jets and \MET \\
		\cline{2-3}
		& $p p \to t b H^\pm~(H^\pm\to W^\pm H)$ & Same as above with extra $b$-jet \\
		\cline{2-3}
		& $pp\to H^\pm H^\mp~(H^\pm\to HW^\pm)$ & $\to6W\to3~\text{same sign leptons}$ + jets and \MET \\
		\cline{2-3}
		& $p p \to H^\pm W^\pm~(H^\pm\to HW^\pm)$ & $\to6W\to3~\text{same sign leptons}$ + jets and \MET \\
		\hline
		\multirow{4}{10pt}{$A$} & $gg\to A$ ($gg$F) & $\to t\bar{t}$ \\
		& & $\to\gamma\gamma$ \\
		\cline{2-3}
		& $gg\to A\to Z H~(H\to SS/Sh)$ & Same as $pp\to ZH$ above, but with resonance structure over final state objects \\
		\cline{2-3}
		& $gg\to A\to W^\pm H^\mp (H^\mp\to W^\mp H)$ & $6W$ signature with resonance structure over final state objects \\
		\hline
		\multirow{2}{10pt}{$S$} & $gg\to S$ ($gg$F) & Resonantly through decays as in Table~\ref{tab:i} ($\gamma\gamma$, $b\bar{b}$, $\tau\tau$, $ZZ\to4\ell$) \\
		& \textit{or} $H\to SS/Sh$ (associated production) & Non-resonantly through multilepton + \MET decays \\	
		\hline
	\end{tabular}
	\caption{A list of potential search channels arising from the addition of the new scalars presented in this paper. This list is by no means complete, but contains clean search channels which could make for striking signatures in the LHC physics regime. Note that in the mass ranges we are considering, $H$ almost always decays to $SS$ or $Sh$, where $S$ and $h$ are likely to decay to $W$s or $b$-jets.}
	\label{tbl:los}
\end{table*}

\subsection{The additional scalars $S$ and $\chi$}
\label{hsx}

The inclusion of $S$ and $\chi$ in the model is especially significant in terms its phenomenology, since the signatures arising from the 2HDM scalars have mostly been addressed in other works already. With this in mind, the combination of the 2HDM with $\chi$ and $S$ can lead to many interesting final states useful for study -- lists of these can be seen in Tables~\ref{tab:i} and \ref{tbl:los}.

The dominant production mechanism of $S$ is assumed to be through the decay processes $H\to SS$ and $H\to Sh$. The admixture of these decays is controlled by a ratio of BRs, defined by $a_1\equiv{ \text{BR}(H\rightarrow SS)  \over \text{BR}(H\rightarrow Sh)}$. $S$ is assumed to be similar to the SM Higgs boson, in the sense that its couplings to SM particles have the same structure as $h$. These couplings are then dependent on $m_S$, and a choice of $m_S$ therefore has implications on the final states that can be studied. Within the mass range considered (i.e. between $m_h$ and $m_H - m_h$), $S$ can be in one of two regions. The first is dominated by $S\to VV$, when $m_S\gtrsim 2m_W\sim 160$~GeV. The second is when $m_S\lesssim 2m_W$, and in this region $S$ has non-negligible BRs to various decay products such as $b\bar{b}$, $VV$, $gg$, $\gamma\gamma$, $Z\gamma$ etc. 

In this model, $S$ is also assumed to be a portal to DM interactions through the decay mode $S\to\chi\chi$. With all other couplings to SM particles fixed, the BR to $\chi\chi$ is a free parameter in the theory. When adding this decay mode, all of the SM decay modes are scaled down by $1-\text{BR}(S\to\chi\chi)$, and the total width of $S$ increases accordingly (although in practical studies, a narrow width approximation will suffice).

The SM Higgs boson has stringent experimental limits on its invisible BR. In this model, this is interpreted by the fact that the $h\to\chi\chi$ BR is suppressed by the choice of $m_\chi\sim m_h/2$. Therefore, $S$ is an important component of the model since is useful to study events which can have an arbitrarily large amount of \MET depending on $m_H$, $m_S$ and $\text{BR}(S\to\chi\chi)$.

%In the effective theory approach we introduced two additional singlet scalars $\chi$ and $S$; this was also done on introducing the modified Type-II
%2HDM which we have constructed. 
%As mentioned, $\chi$ is a scalar DM candidate which allows us to study signatures relating to \MET.
%
%A few phenomenological aspects of the scalar $S$ have already been discussed while considering
%the effective theory approach. By introducing it as a Higgs boson-like particle with a narrow mass range and various decay
%possibilities (as shown in Table~\ref{tab:i}), we are open to interesting final states for study. By making the choice that $S$ is Higgs-like, the number of free parameters in the theory is drastically reduced in the sense that all of it's decay modes have fixed BRs. Therefore, with $S$ having a possibility to decay through $S\to VV$, a large variety of leptonic signatures can be explored. A few of these will be discussed in the next section.
  
\section{Analysis of selected leptonic signatures}
\label{anres}

In order to understand the impact that the model has on certain leptonic final states, a series of analyses are presented in this section. For these, we consider the following mass ranges for each new particle:
\begin{itemize}
\item [(a)] Light Higgs: $m_h = 125$~GeV (\text{assumed as the SM Higgs}),
\item [(b)] Heavy Higgs: $2 m_h < m_H < 2 m_t$,
\item [(c)] $CP$-odd Higgs: $m_A > \left(m_H + m_Z\right)$,
\item [(d)] Charged Higgs: $\left(m_H + m_W\right) < m_{H^\pm} < m_A$,
\item [(e)] Additional scalars: $m_\chi < m_h/2$ and $m_h \lesssim m_S \lesssim \left(m_H - m_h\right)$.  
\end{itemize}

Based on these mass choices, we can study the BRs of 2HDM scalars into the SM particles and the additional scalars
$\chi$ and $S$ as listed in Table~\ref{tab:i} with following production channels:
\begin{itemize}
\item [(a)] $g g \to h$, $H$, $A$, $S$,
\item [(b)] $p p \to t H^- (\bar t H^+)$, $t H^-\bar b + \bar t H^+ b$, $H^+ H^-$, $H^\pm W^\pm$. 
\end{itemize}

There are many interesting phenomenological aspects we can consider with the combination of production and decay modes 
discussed above (the theory pertaining to the dominant production modes of $A, H$ through $gg$F in a Type-II 2HDM are given in \ref{appenprod} and list of several search modes are listed in Table~\ref{tbl:los}). 
It is not feasible to analyse all of the possible final states that could provide potential for discovery. As a case study, we rather focus on a few striking signatures driven by the production of multiple leptons. These signatures are also dependent on the production of a non-negligible amount of \MET. However, the signatures have been chosen such that the first two (Sects.~\ref{sec:4l} and \ref{sec:3lSS}) do not rely on the $S$'s interaction with DM, whereas the third (Sect.~\ref{sec:AZH}) does. This is an example of how the ``simplified model'' approach is useful in that different searches can be used to constrain different parameters of the theory.

For Sects.~\ref{sec:4l}, \ref{sec:3lSS} and \ref{sec:AZH}, some plots of key signature distributions are shown and discussed. These plots were made from selecting Monte Carlo (MC) events generated in \textsc{Pythia 8.219}~\cite{Sjostrand:2014zea} using custom \textsc{Rivet}~\cite{Buckley:2010ar} routines. In all three cases, 500~000 events were generated and a selection efficiency was determined based on cuts and criteria. These events are not passed through a detector simulation. The reason for this is that our intentions are not to model the profile of \MET with accuracy, but rather provide a signature of the general region in which \MET could be expected, given the parameter constraints.\footnote{Having said this, the state of the art fast simulation package \textsc{Delphes 3}'s~\cite{deFavereau:2013fsa} predictions of detector effects in \MET are reasonable, but still not completely compatible with the full simulation packages used by ATLAS and CMS. Detector simulation could be studied in a future work.} For the first two analyses, leptons were defined as either electrons or muons with $p_\text{T}>15$~GeV and $|\eta|<2.47~(2.7)$ for electrons (muons). A crude lepton isolation is applied by vetoing any leptons which share a partner lepton within a cone of radius $\Delta R=\sqrt{(\Delta\phi)^2+(\Delta\eta)^2}=0.2$ around it, and any leptons coming from a hadron decay are vetoed.

The mass points considered in these distributions are relatively close to the central points in the ranges we are considering here. The mass of $S$ is fixed to 150 GeV, where it still enjoys a wide range of decay modes due to it's SM-like nature -- at this mass the BRs to $b\bar{b}$ and $VV$ are both non-negligible allowing for sensitivity in di-jet and di-boson searches, while a lighter $S$ runs the risk of being too close to the Higgs mass for a comfortable experimental resolution. The mass of $H$ is considered at the two values 275 GeV and 300 GeV. A mass close to 275 GeV does have some motivation from Ref.~\cite{vonBuddenbrock:2015ema}, but is also interesting since the $H\to SS$ decay is then off-shell. The on-shell behaviour is probed by also selecting the point $m_H=300$~GeV, and $a_1$ is used is chosen such that $\text{BR}(H\to SS)=\text{BR}(H\to Sh)=0.5$ in order for both decay mechanisms to be explored evenly. $\text{BR}(S\to \chi\chi)$ is chosen to be $0.5$ to probe intermediate \MET production mechanisms.

%for example, if we consider $H^\pm$ production in association with $t$-quark or $W^\pm$ bosons
%where $t$-quark and $W^\pm$ decay into leptonic final states then we have very clean signature of observing same-sign
%multi-lepton final states with less background. Further interesting channels of interest are described in following sections.

\subsection{$H\rightarrow 4W\rightarrow  4l+\MET$}
\label{sec:4l}

\begin{figure*}
\centering 
\includegraphics[height=.40\textwidth]{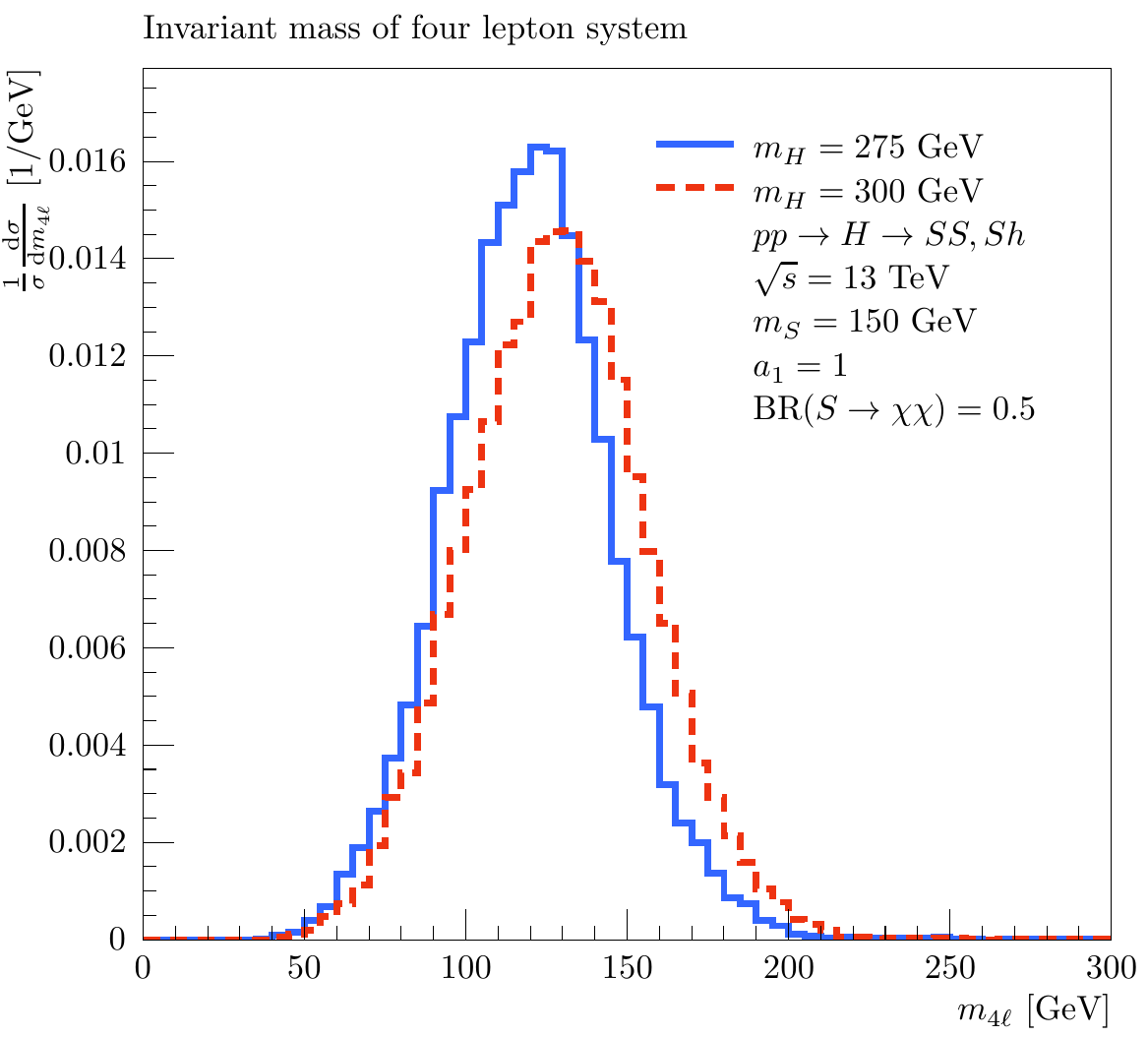}
\includegraphics[height=.40\textwidth]{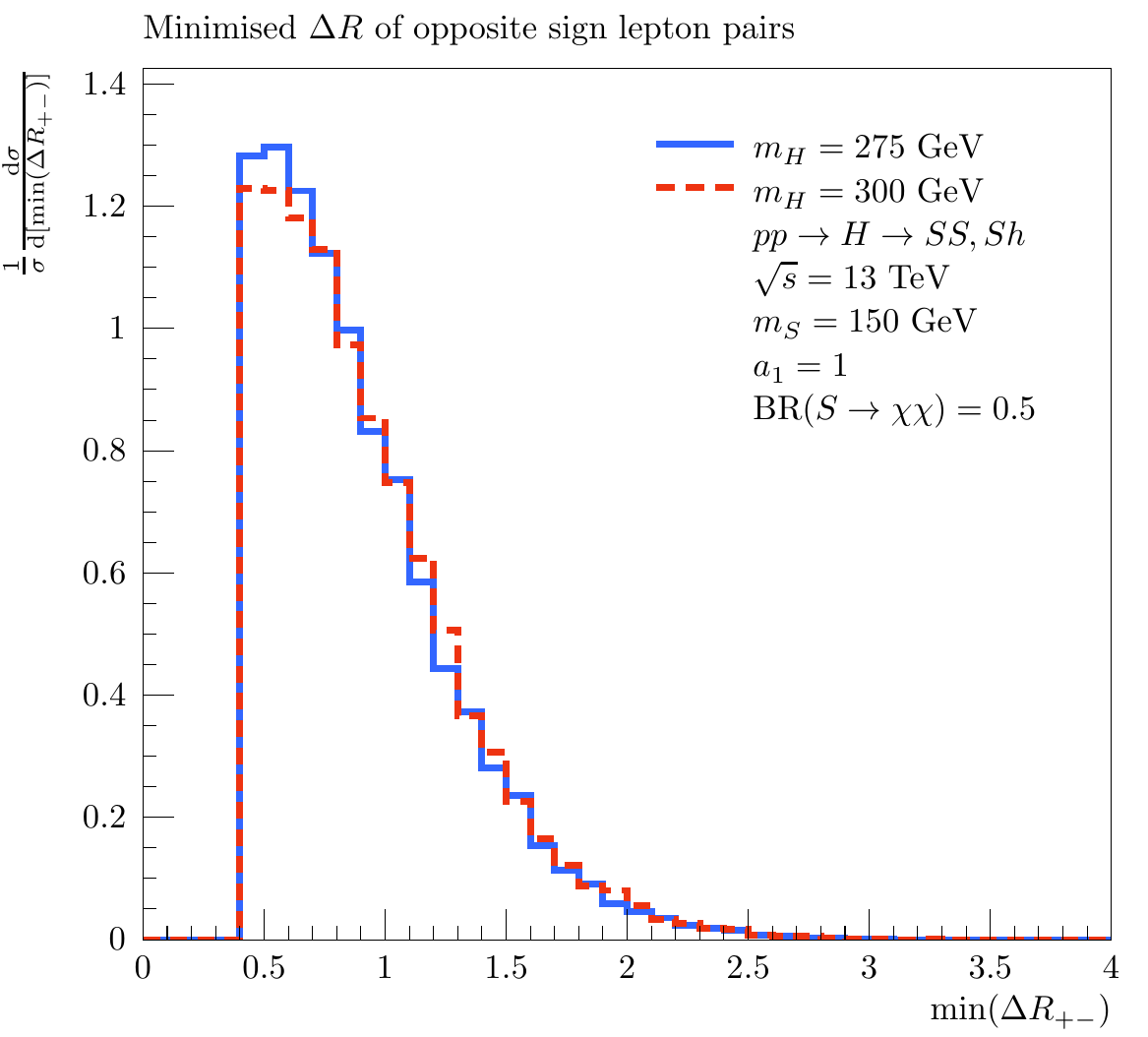}

\includegraphics[height=.40\textwidth]{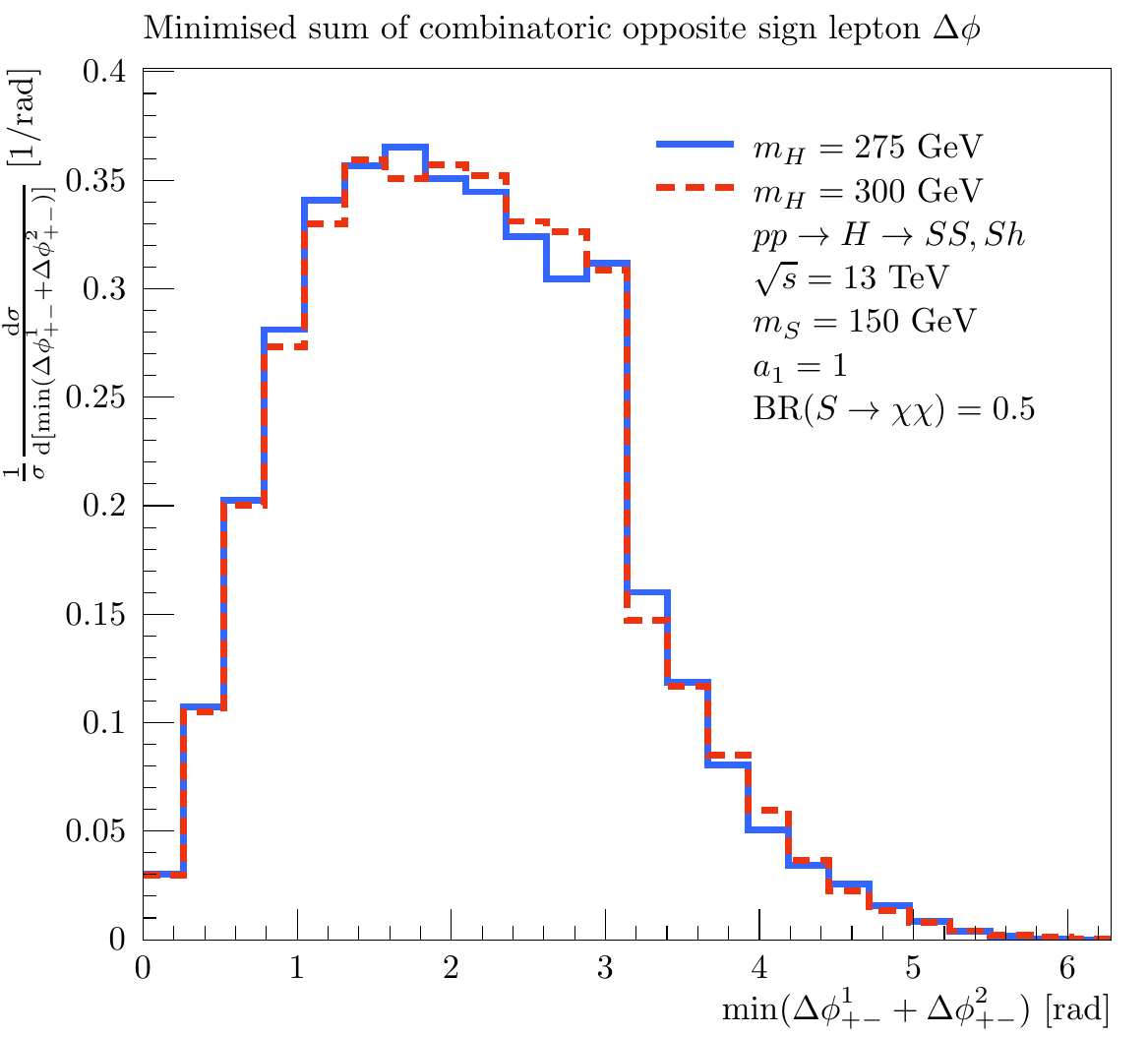}
\includegraphics[height=.40\textwidth]{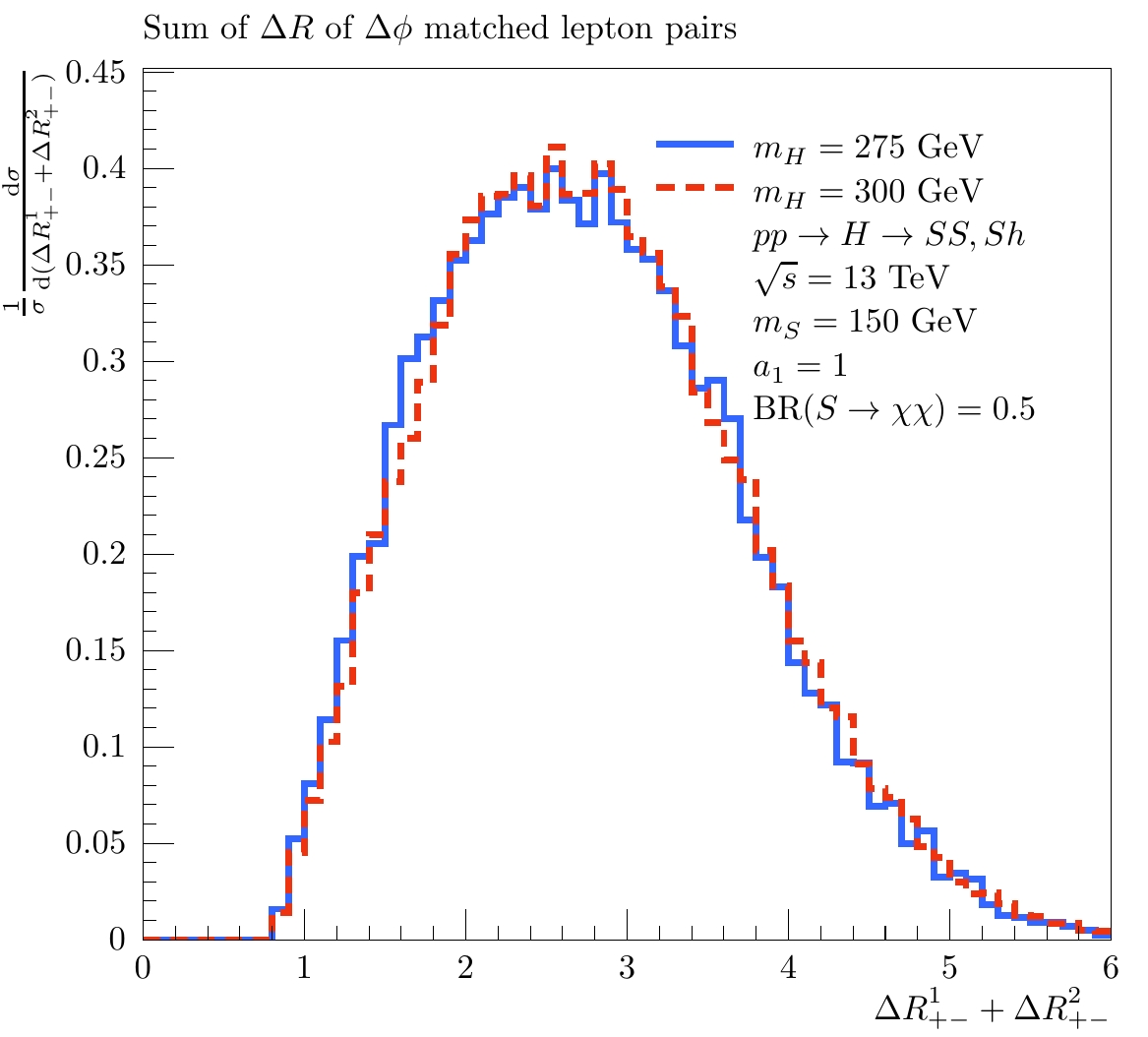}

\includegraphics[height=.40\textwidth]{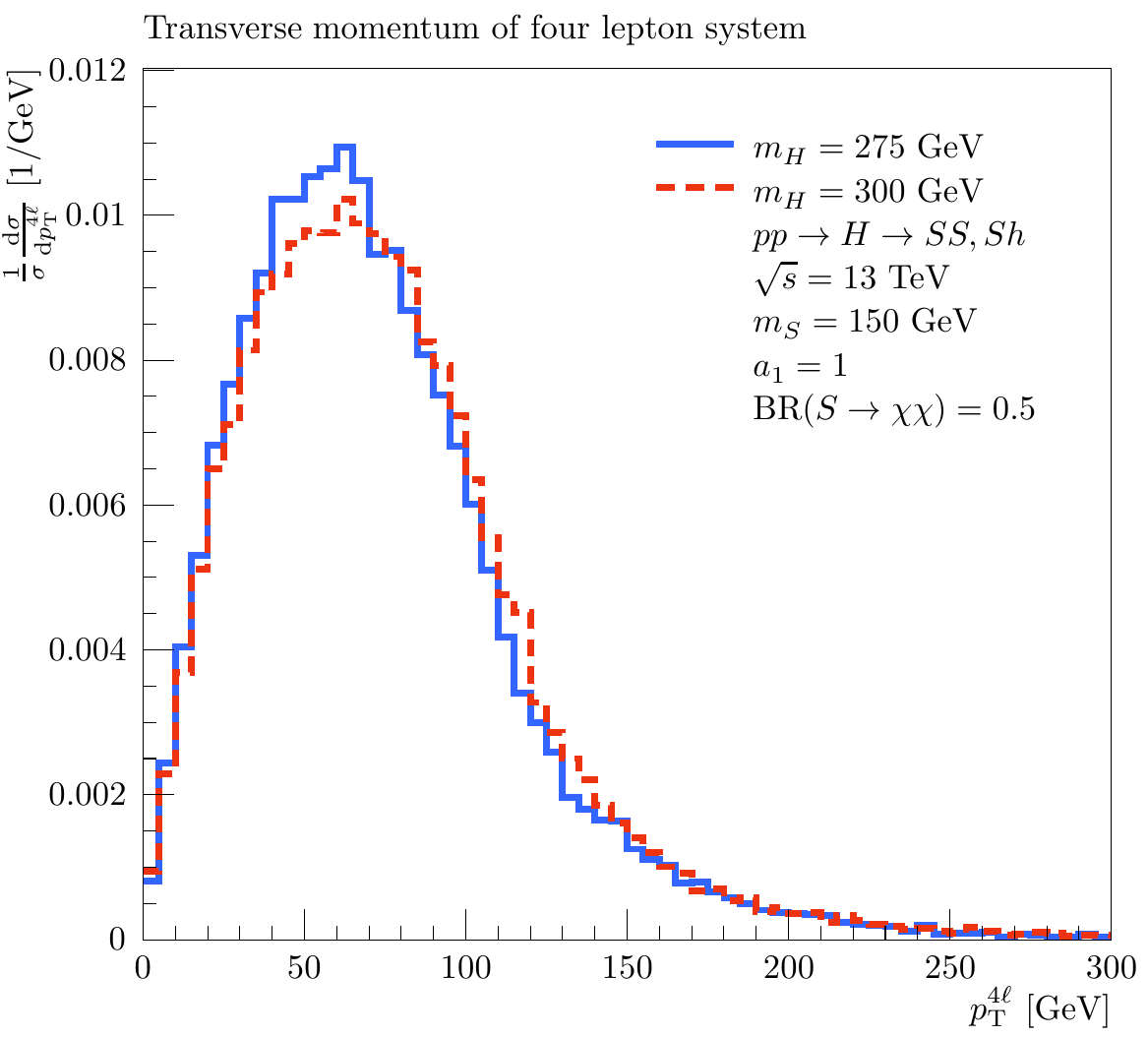}
\includegraphics[height=.40\textwidth]{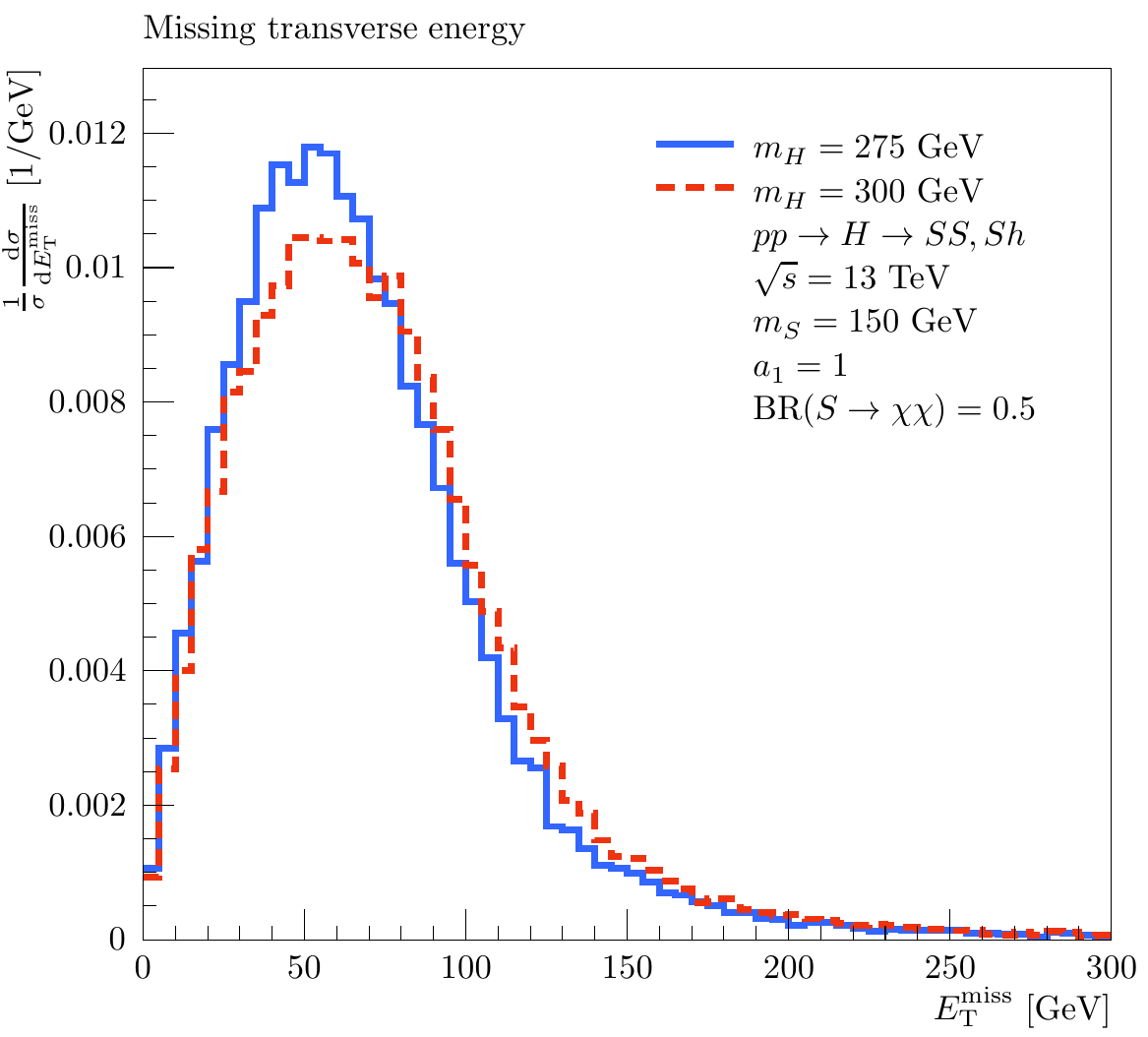}
\caption{\label{fig:4l} Various leptonic kinematic distributions (normalised to unity) pertaining to the process $H\rightarrow 4W\rightarrow  4l+\MET$, as described in Sect.~\ref{sec:4l}. }
\end{figure*}

Assuming a large enough cross section for the single production of $H$, the decays $H\rightarrow SS, Sh$ can lead to a sizeable production of 4 $W$s. The leptonic decays would produce 4 charged leptons ($e, \mu$) in conjunction with large \MET. Due to the spin-0 nature of the $S, h$ bosons the leptons of the decay of each boson appear close together~\cite{Dittmar:1996ss}, leading to an even more striking signature. 

Fig.~\ref{fig:4l} displays the kinematics of the leptons for $m_H=275, 300\,$GeV and $m_S=150\,$GeV for a proton-proton centre of mass energy of 13~TeV. Results are shown assuming  $a_1=1$ and $\text{BR}(S\rightarrow\chi\chi)=0.5$. In the event generation, both $S$ and $h$ are forced to decay to $WW$, and these $W$s are forced to decay semi-leptonically (including $\tau\nu_\tau$ decays, since these can result in final states containing muons or electrons). Given the $gg\to H$ cross section range mentioned in Sect.~\ref{hsh}, one could expect a cross section times BR of as much as about 50~fb for this process at the mass points considered here.

The upper left plot shows the invariant mass of the 4-lepton system ($m_{4l}$). In the mass range of interest here the background is suppressed and it is dominated by the non-resonant production of di-$Z$ bosons in which at least one is off-shell~\cite{Chatrchyan:2013mxa,Aad:2014eva}. The production of the SM Higgs boson would need to be taken into account as a background. The contribution from processes where at least one lepton arises from hadronic decays is sub-leading to the production of $pp\rightarrow ZZ^*\rightarrow 4l$.

The upper right plot displays a distribution of the smallest $\Delta R$ between opposite sign leptons. This variable exploits the spin-0 nature of the $S,h$ bosons.\footnote{The kinematics of the decay depend on the tensor structure of the $SVV$ coupling.} The distribution suffers from a cut-off due to the requirement that leptons be apart from each other by $\Delta R> 0.4$ due to isolation requirements. The left plot in the middle displays the sum of the di-lepton azimuthal angle separation for the two opposite sign pairs ($\Delta\phi_{+-}$). Here the choice of lepton pairs is performed so as to minimize the sum of the di-lepton azimuthal angle separation. The corresponding sum of $\Delta R$ distances for this choice of lepton pairing is shown in the middle right plot. The lower plot displays the transverse momentum of the 4-lepton system and the \MET. These distributions are significantly different from what one would expect from the residual backgrounds from $pp\rightarrow ZZ^*\rightarrow 4l$.

 The production of  $t\overline{t}Z$ is a source of four charged leptons~\cite{Lazopoulos:2008de}. This background can be suppressed by a combination of requirements including vetoing on the presence of jets and $b$-jets. The production 4$W$s in the standard model is dominated by $t\overline{t}t\overline{t}$~\cite{Bevilacqua:2012em,Alwall:2014hca} and $t\overline{t}WW$~\cite{Alwall:2014hca} are significantly smaller and can be neglected. The production of $t\overline{t}t\overline{t}$ with other final states has been investigated and no significant excess in the data has been observed with respect to the SM prediction~\cite{ATLAS-CONF-2016-020}.

\subsection{$t(t)H\rightarrow 6W\rightarrow  l^{\pm}l^{\pm}l^{\pm}+X$}
\label{sec:3lSS} 

\begin{figure*}
\centering 
\includegraphics[height=.40\textwidth]{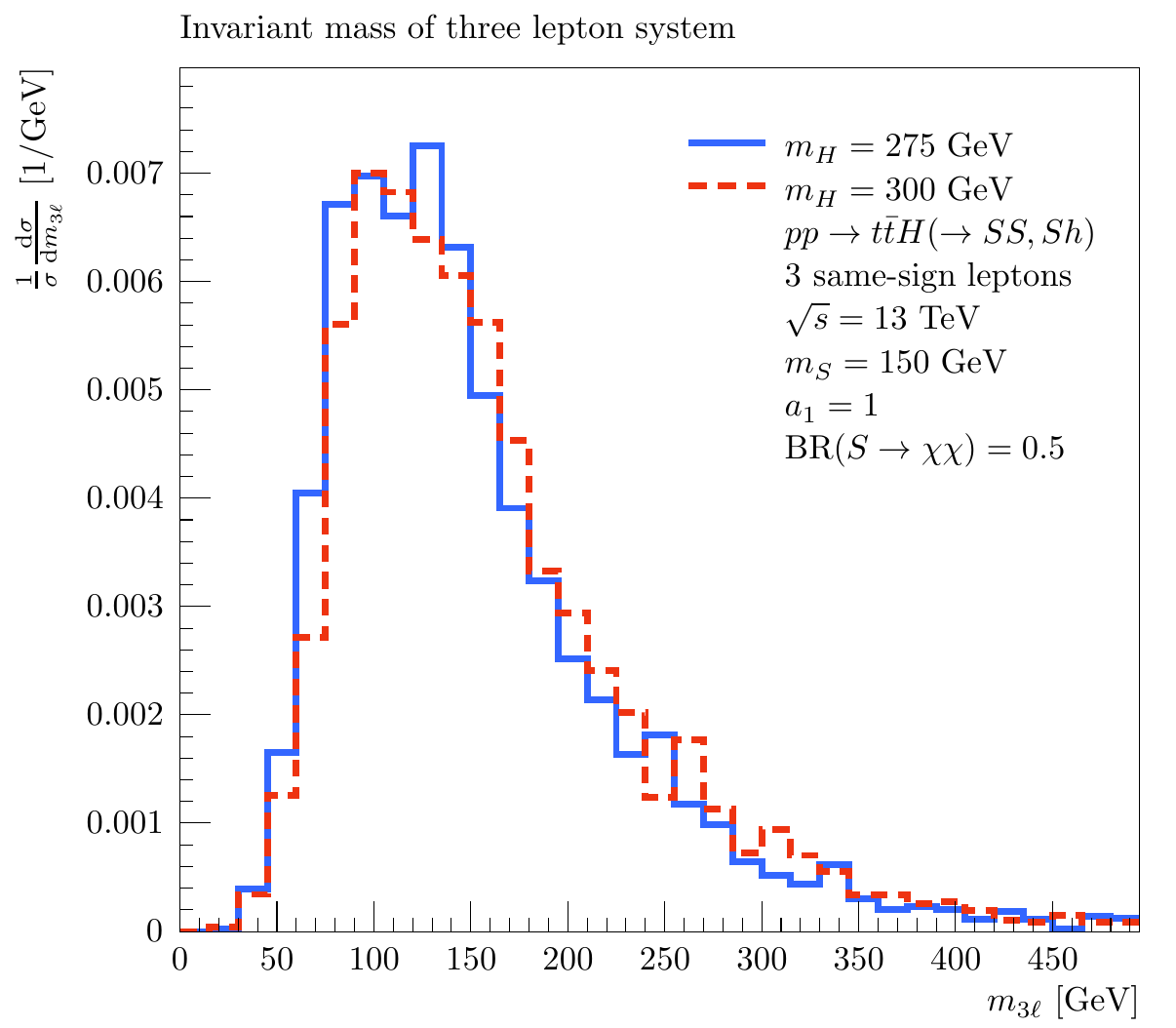}
\includegraphics[height=.40\textwidth]{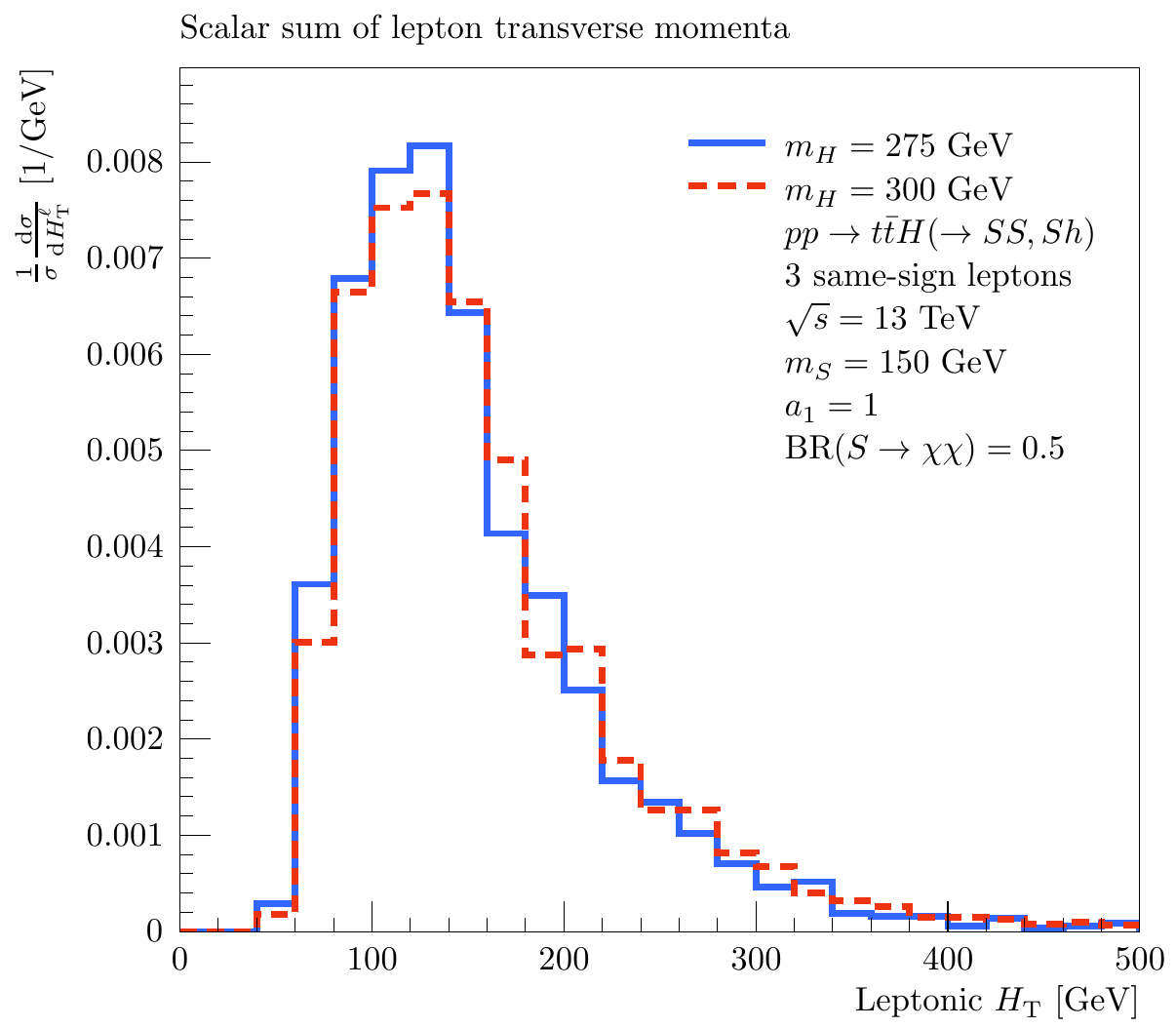}

\includegraphics[height=.40\textwidth]{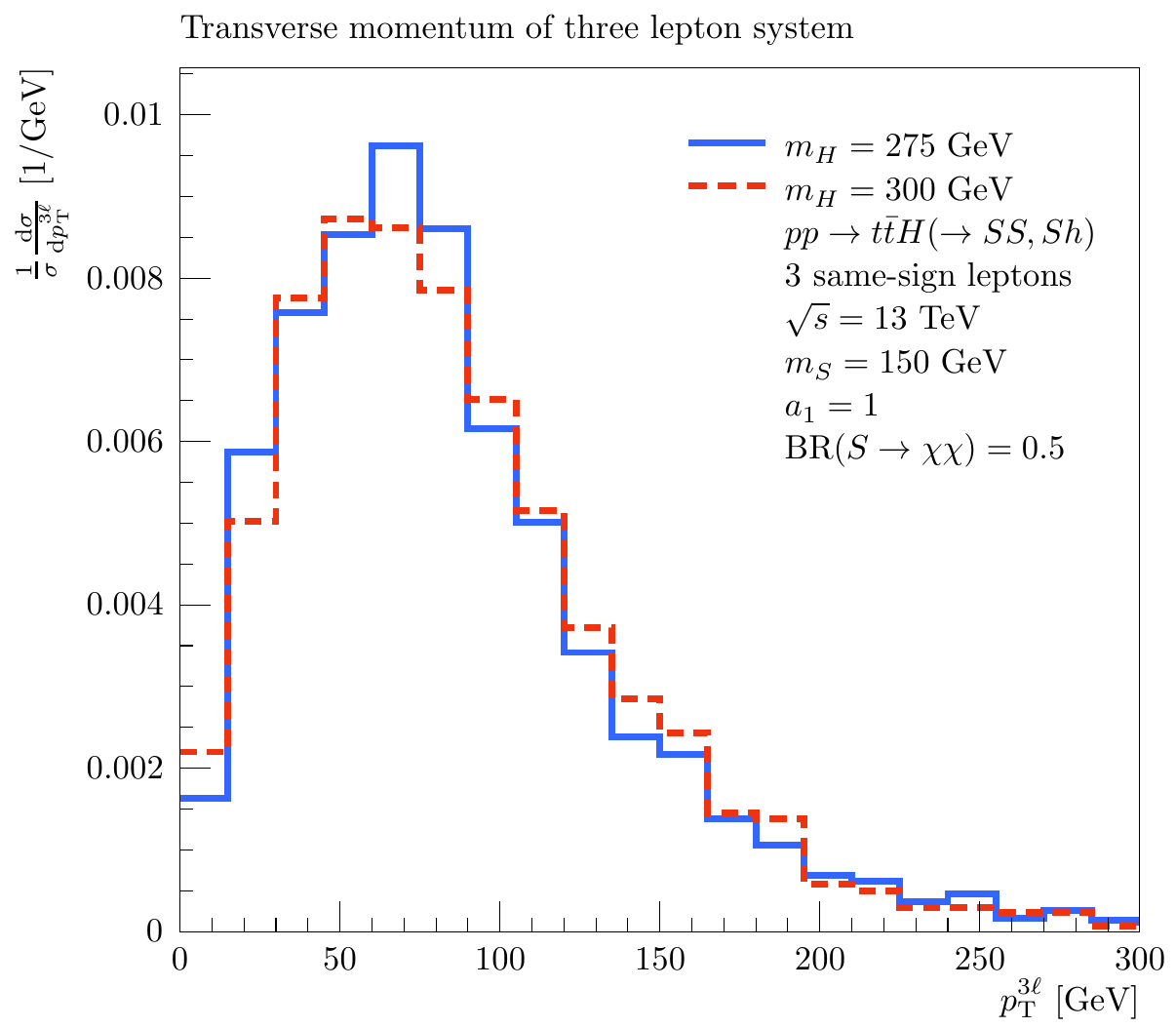}
\includegraphics[height=.40\textwidth]{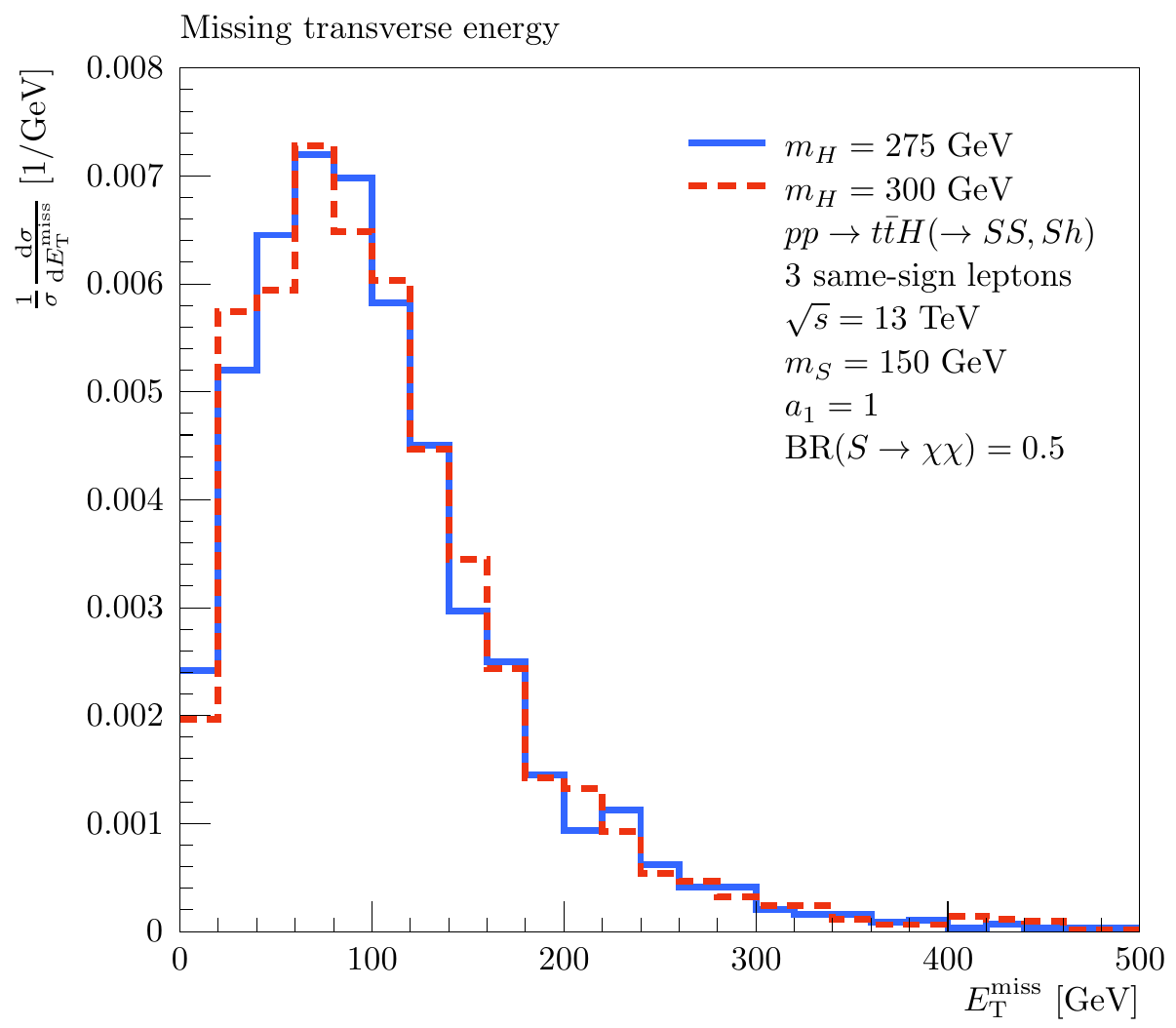}

\includegraphics[height=.40\textwidth]{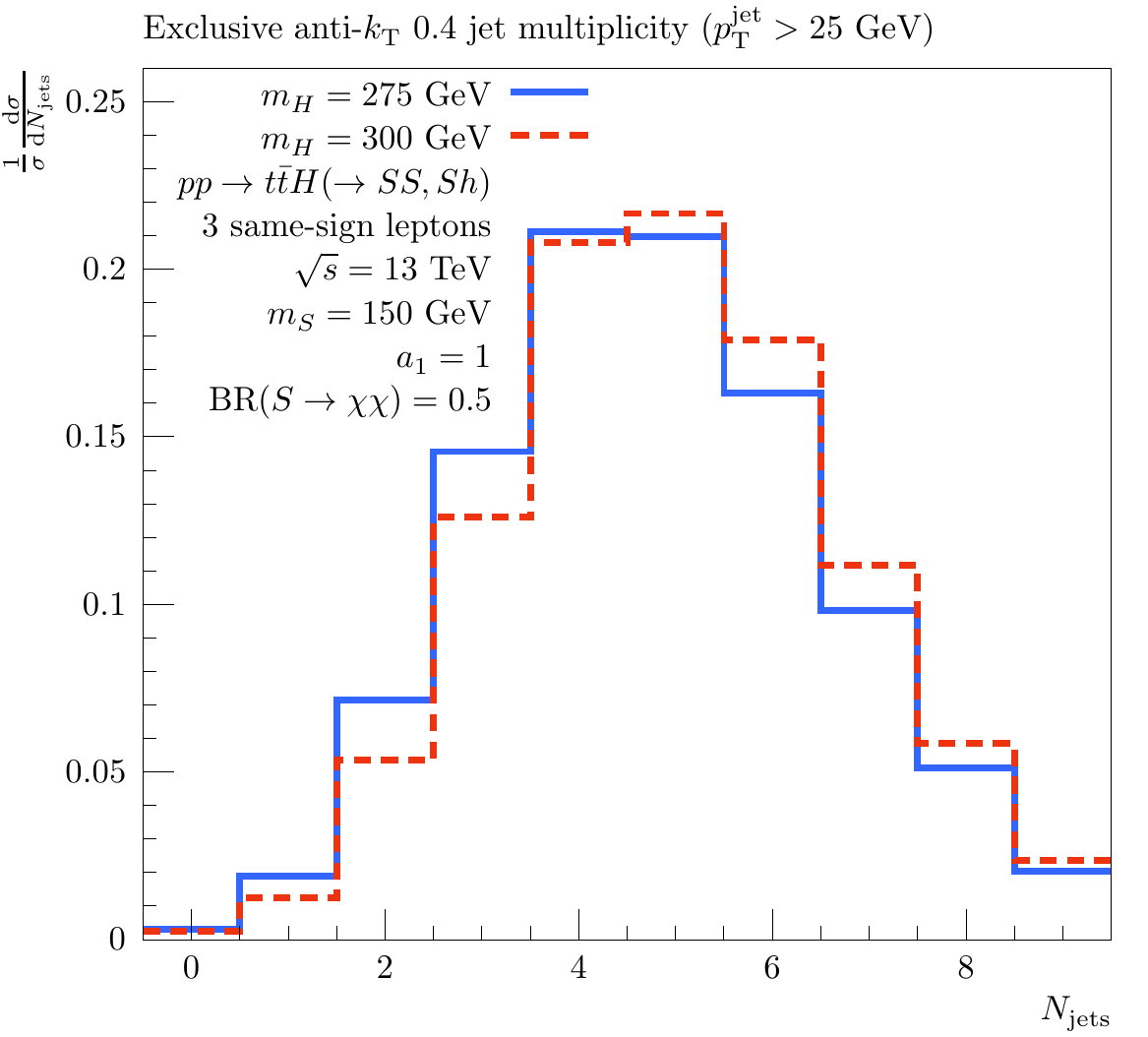}
\includegraphics[height=.40\textwidth]{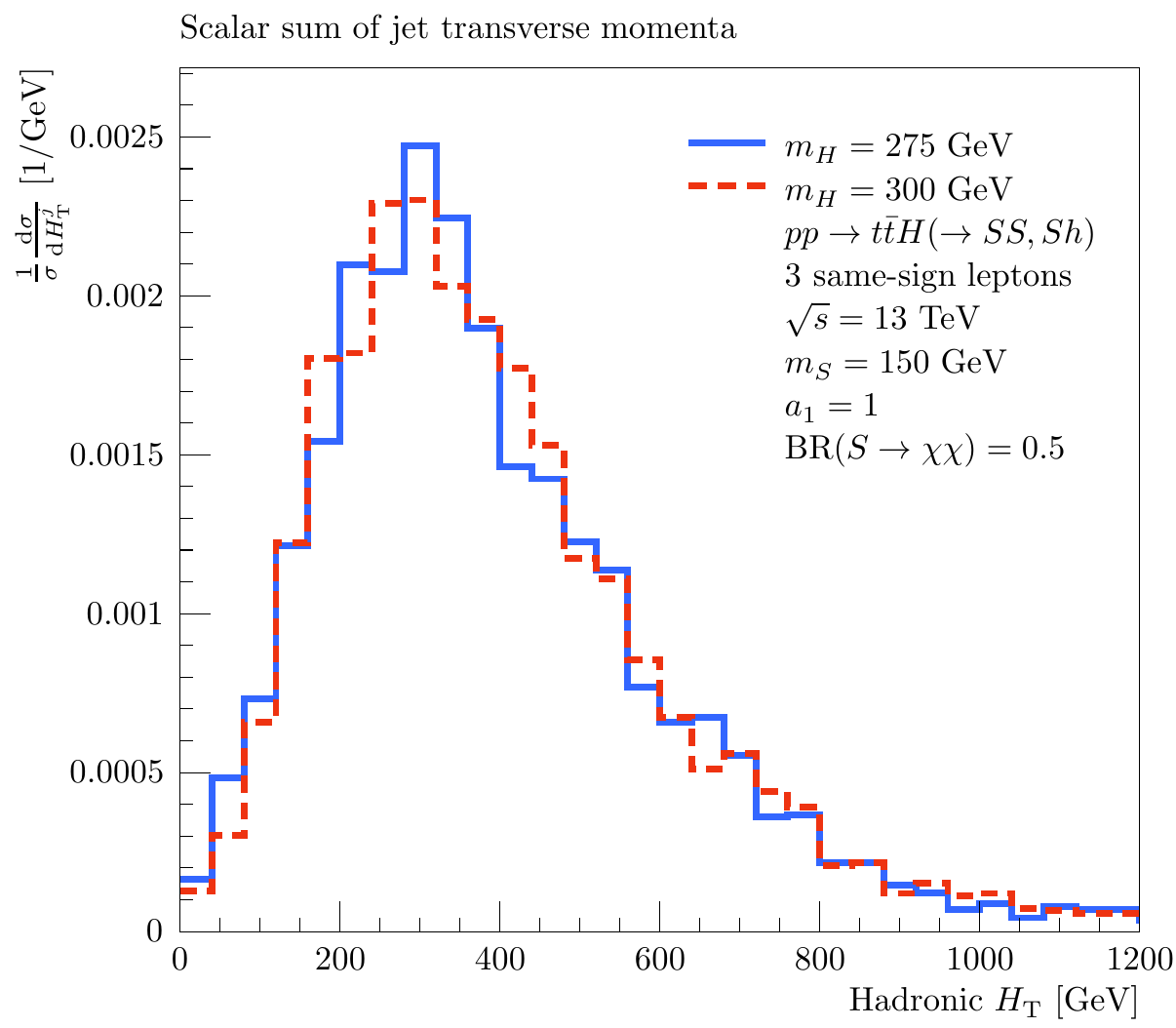}

\caption{\label{fig:3lss} Various hadronic and leptonic kinematic distributions (normalised to unity) pertaining to the process $t\overline{t}H\rightarrow 6W\rightarrow  l^{\pm}l^{\pm}l^{\pm}+X$, as described in Sect.~\ref{sec:3lSS}.}
\end{figure*}

The production of double and single top quarks in association with the heavy scalar produce up to 6 $W$s in association with $b$-quarks. This leads to the possibility of producing three same-sign isolated charged leptons ($l^{\pm}l^{\pm}l^{\pm}$), a unique  signature at hadron colliders. The production of same-sign tri-leptons, including non-isolated leptons from heavy quark decays was, suggested in Ref.~\cite{Barger:1984rp} to tag top events.  The production of isolated same-sign tri-leptons has been studied in the context of the search for new leptons~\cite{Ozcan:2009qm} and in $R$-parity violating SUSY scenarios~\cite{Mukhopadhyaya:2010qf,Mukhopadhyay:2011xs}. Background studies performed in Refs.~\cite{Ozcan:2009qm,Mukhopadhyay:2011xs} indicate that the production of three same-sign isolated leptons is very small, less than $1\times 10^{-2}\,$fb for a proton-proton centre of mass of $13\,$TeV. The background would be dominated by the production of $t\overline{t}W$ with additional leptons from heavy flavour decays. This background is reducible by means of isolation, impact parameters and other requirements~\cite{Chatrchyan:2013mxa,Aad:2014eva}. With a reasonable choice of parameters a fiducial cross section of $0.5$~fb can be predicted for $13$~TeV centre of mass energy, rendering the search effectively background free.

It is relevant to study the kinematics of the final state here, as detailed in Fig.~\ref{fig:3lss}. The event generation allowed for the decay of $S$ and $h$ into any channels involving a $W$, $Z$ or $\tau$. To ensure a clean signal, leptons were only selected if they did not come from a hadron decay -- these processes contain many $B$-hadrons which can decay into leptons. Under these conditions, the efficiency in selected at least 3 leptons in an event was about 8\%. Of these events, about 15\% would contain a group of three same-sign leptons. The upper left and right plots display tri-lepton invariant mass and the scalar sum of the transverse momenta ($H_\text{T}$) of the leptons, respectively. The transverse momentum of the three leptons is shown in the middle left plot. The \MET distribution is shown in the middle right plot. The average \MET in these evens is significant and it adds to the uniqueness of the signature.

Since the production of three same-sign isolated leptons requires the presence of at least six weak bosons and/or $\tau$ leptons, a large number of jets is expected from those particles that do not decay leptonically. This makes the production of three same-sign isolated leptons even more striking. Hadronic jets are defined using the anti-$k_T$ algorithm~\cite{Cacciari:2008gp} with the parameter $R=0.4$. Jets are required to have transverse momentum $p_\text{T}>25\,$GeV and to be in the range $\left|\eta\right|<2.5$. The jet multiplicity of jets is shown in the lower left plot. The distribution peaks around 4-5 with a long tail stretching to 8 or more jets. The differences displayed by changing $m_H$ are due to the fact that in the case of $m_H=275\,$GeV one of the $S$ bosons in $H\rightarrow SS$ becomes off-shell, reducing the transverse momentum of the jets. The $H_\text{T}$ constructed with jets is shown in the lower right plot.

It is worth noting that the distributions shown in Fig.~\ref{fig:3lss}  also apply to the combination of three leptons where the total charge is $\pm 1$. There the SM backgrounds are significant, although the signal rate is about 6 times larger. 

The production of $H$ with single top is not suppressed with respect to the $t\overline{t}$ production, as it is in the production of the SM Higgs boson. The kinematic distributions shown in Fig.~\ref{fig:3lss} are similar to those displayed by the $tH$ production with the exception of the net multiplicity and the jet $H_\text{T}$, due to the reduced production of $b$-jets. Similar discussion applies to the production of $H^{\pm}\rightarrow W^{\pm}H$.

\subsection{$A\to ZH\to Z+\text{jets}+\MET$}
\label{sec:AZH}

\begin{figure*}
	\centering 
	\includegraphics[height=.40\textwidth]{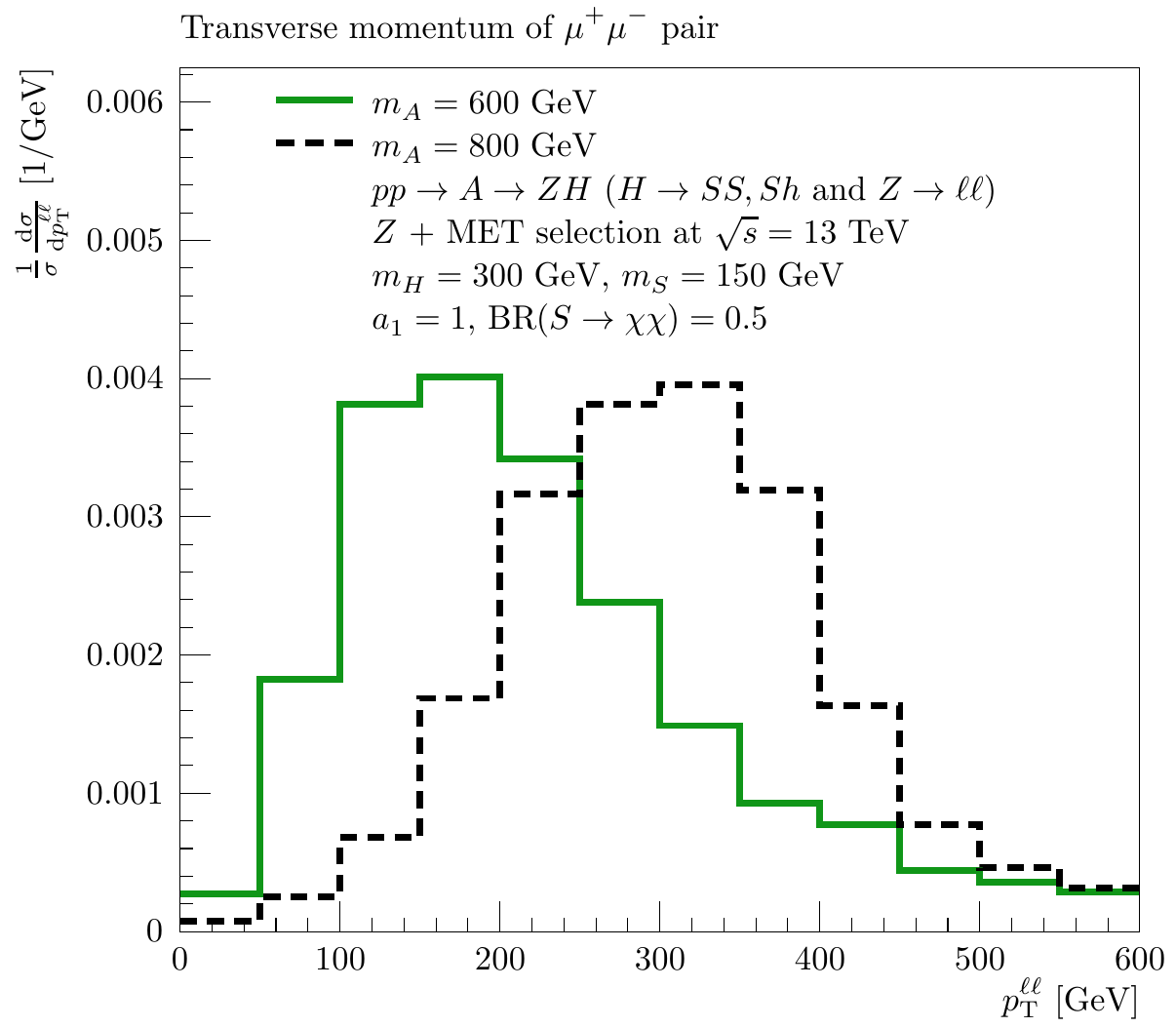}
	\includegraphics[height=.40\textwidth]{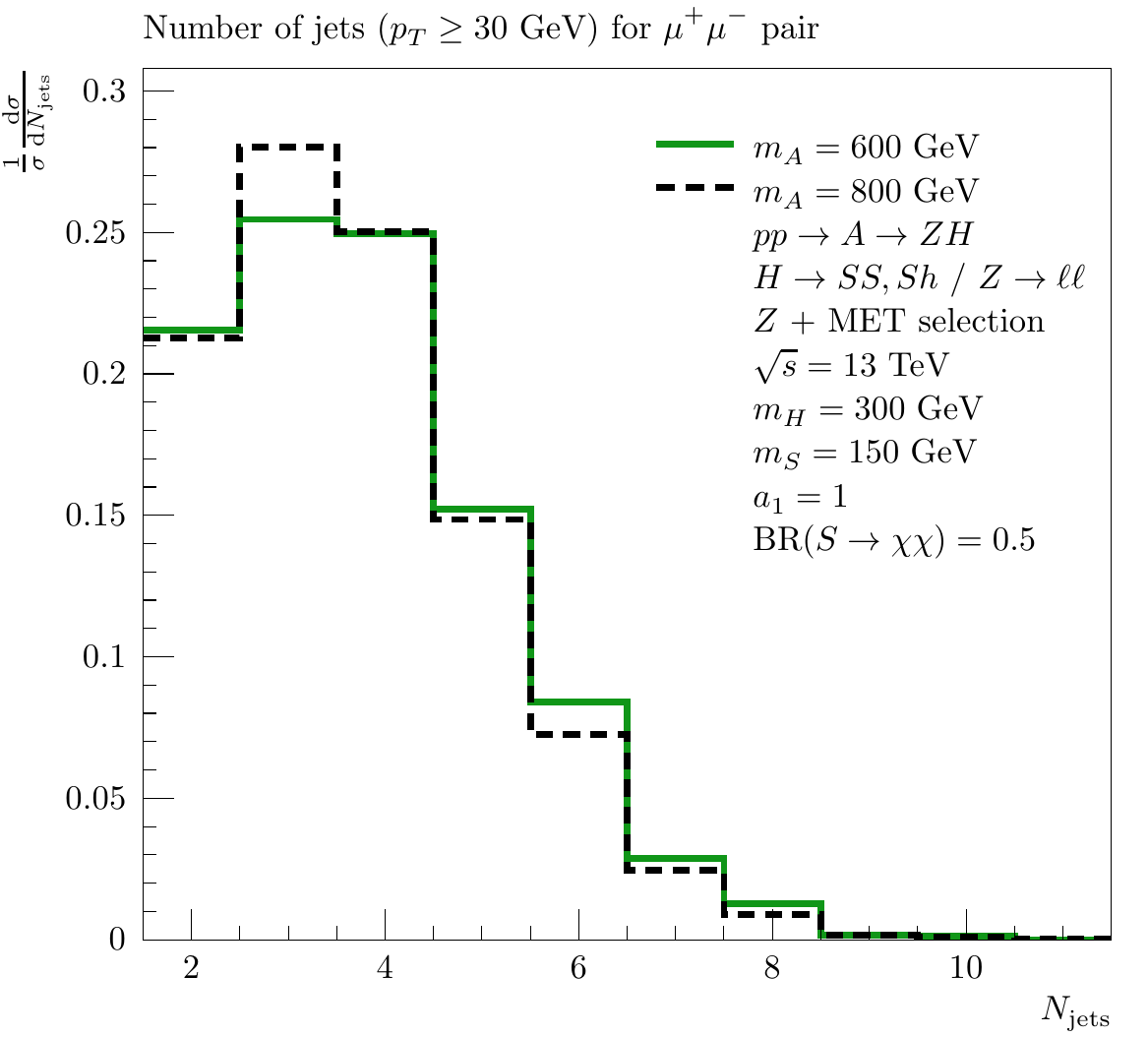}
	
	\includegraphics[height=.40\textwidth]{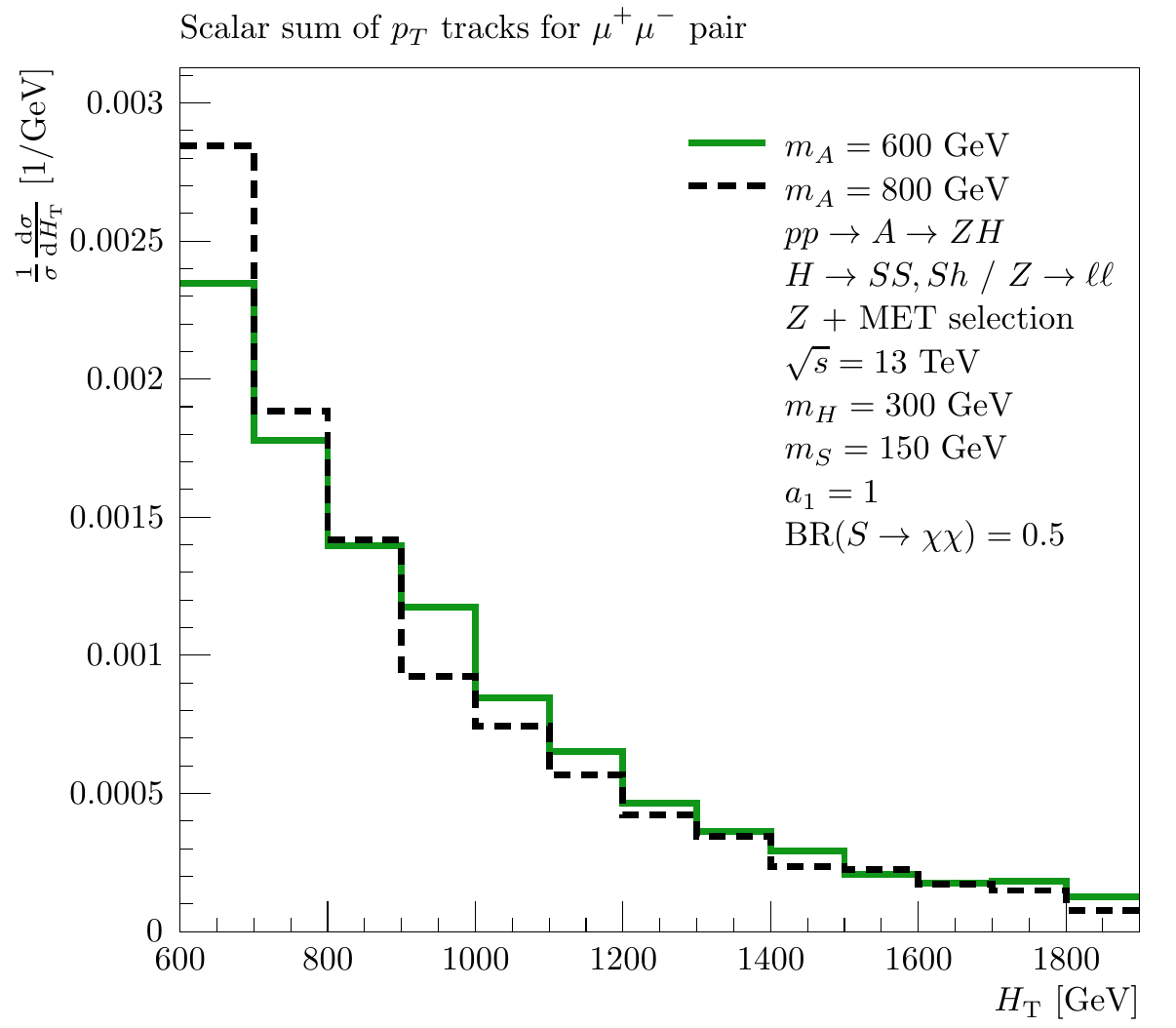}
	\includegraphics[height=.40\textwidth]{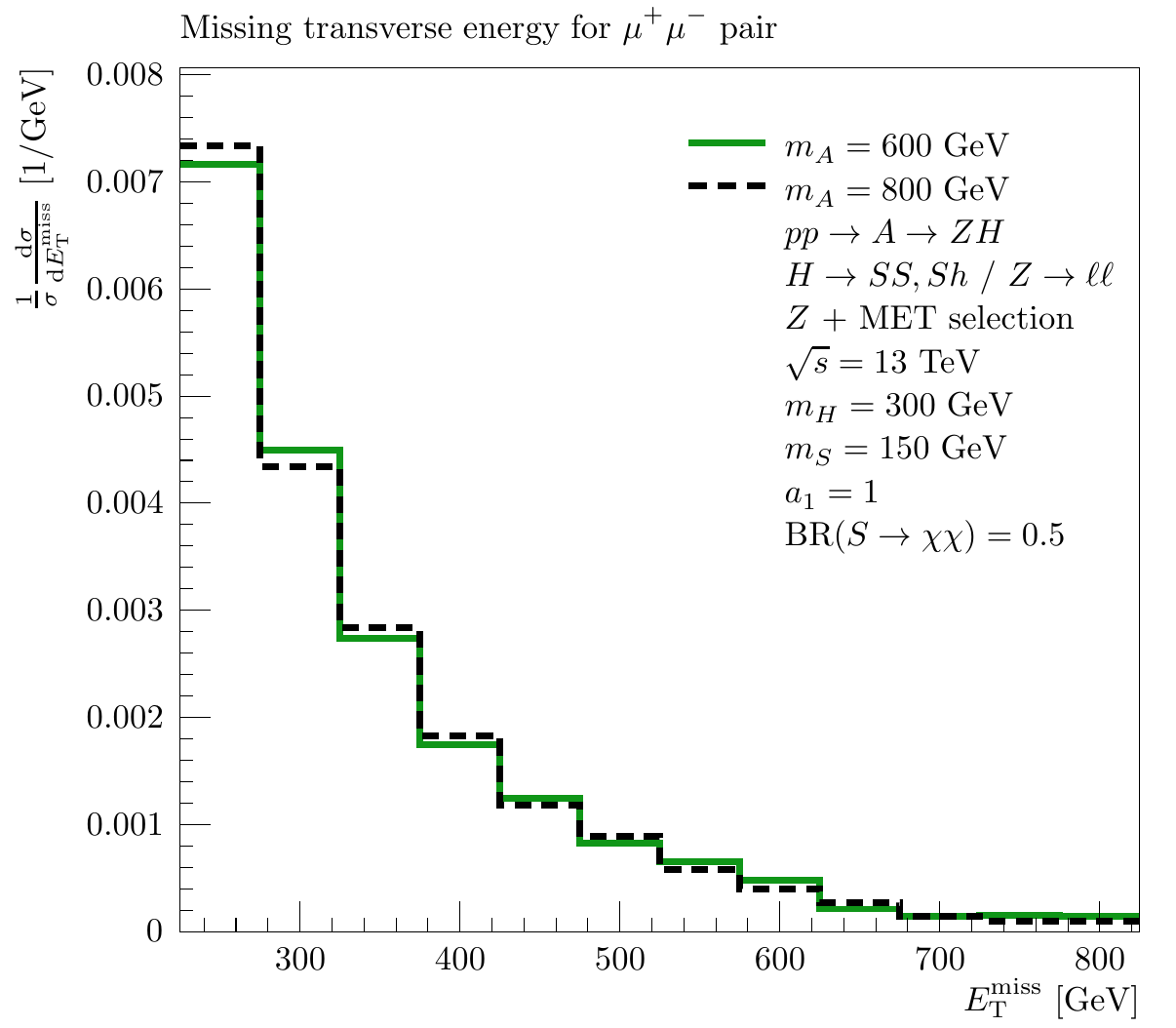}
	
	\includegraphics[height=.40\textwidth]{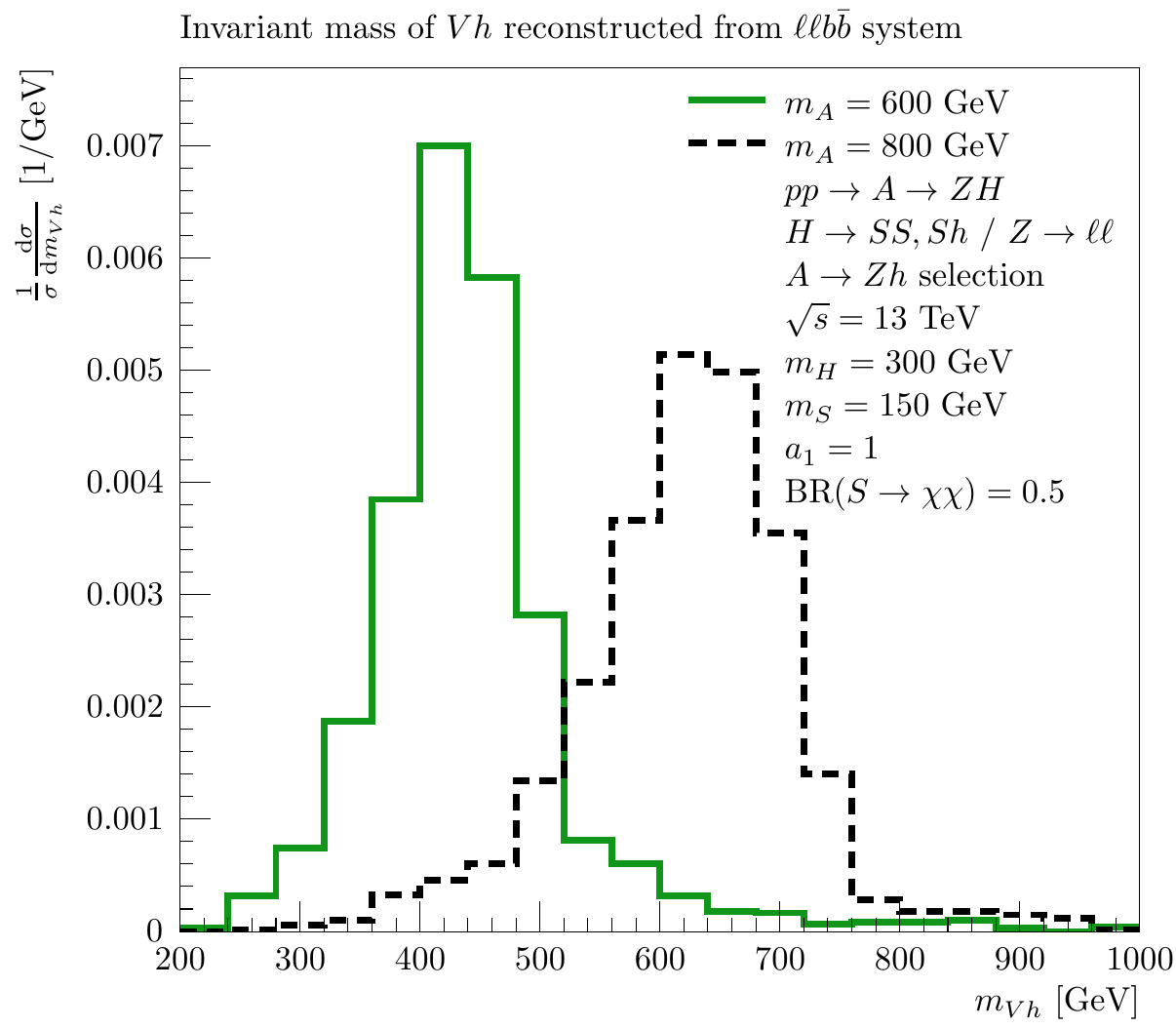}
	\includegraphics[height=.40\textwidth]{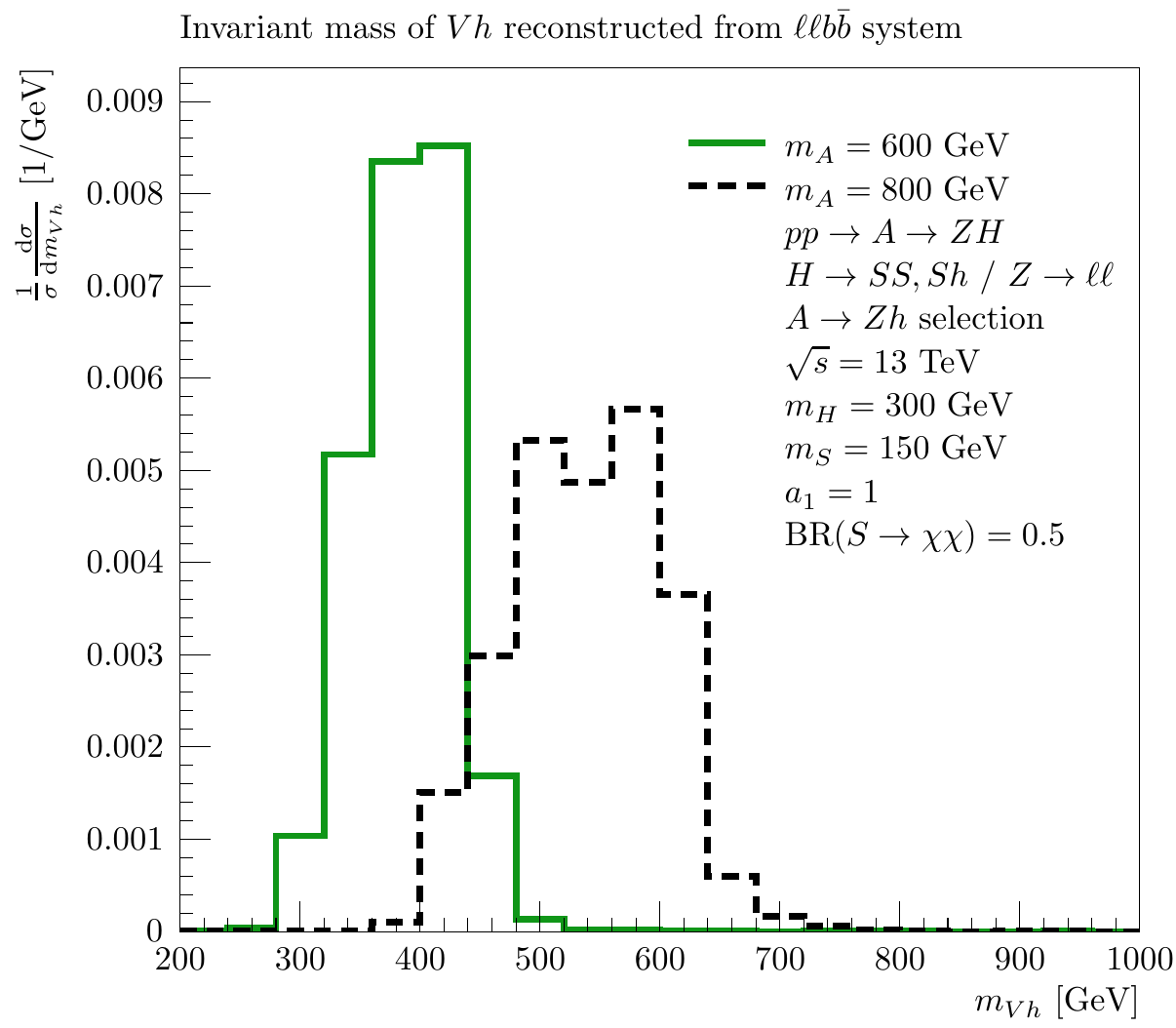}
	\caption{\label{fig:AZH} Kinematic distributions of the leptons in $A\to ZH$, where $H\to SS,~Sh$. The top four pertain to the ATLAS Run 2 $Z+\MET$ SR-Z selection, where the $\mu\mu$ properties are studied since its efficiency is slightly higher than that of $ee$. The bottom two figures pertain to the ATLAS Run 2 $A\to Zh~(h\to b\bar{b})$ selection, with the 1 $b$-tag category on the left and the 2 $b$-tag category on the right.}
\end{figure*}

If we consider Eq.~\ref{eqn:vphiphi}, we note that in the limit where $\cos(\beta-\alpha)\to 0$ (and therefore $\sin(\beta-\alpha)\to 1$), the coupling strength in $A$-$Z$-$H$ becomes large -- this limit applies in the case where $H$ is SM-like. For this reason, a prime search channel for $A$ lies in the $A\to ZH$ decay, if $m_A$ is large enough. If $H\to SS,~Sh$, then there are two obvious LHC based searches which could already shed light on this decay mode. These are the typical SUSY $Z+\MET$~\cite{ATLAS_ZMET_RUN2,Aad:2015wqa,Khachatryan:2015lwa} and the $Zh$ (where $h\to b\bar{b},~\tau\tau$) searches~\cite{Aad:2015wra,ATLAS_Zh_RUN2,Khachatryan:2015lba}.

\begin{table}[t]
	\centering
	\begin{tabular}{c|c|c}
		\hline
		\textbf{Channel/region} & \textbf{Prediction} & \textbf{Experimental limit} \\ 
		\hline
		\multicolumn{3}{c}{\emph{Monojet with $gg\to H\to SS\to 4\chi$ at $\sqrt{s}=8$~TeV}}  \\
		\hline
		$\MET>250$~GeV & $15.1\pm0.18$~fb  & 229~fb \\ 
		$>300$~GeV & $8.90\pm0.063$~fb & 98.5~fb  \\
		$>350$~GeV & $5.42\pm0.023$~fb & 48.8~fb \\
		$>400$~GeV & $3.42\pm0.0093$~fb & 20.2~fb \\
		$>450$~GeV & $2.24\pm0.0040$~fb & 7.82~fb \\
		$>500$~GeV & $1.48\pm0.0017$~fb & 6.09~fb\\
		$>550$~GeV & $1.00\pm0.00080$~fb & 7.21~fb \\
		\hline
		\multicolumn{3}{c}{\emph{$b\bar{b}+\MET$ with $gg\to H\to Sh\to b\bar{b}\chi\chi$ at $\sqrt{s}=13$~TeV}}  \\
		\hline
		Signal region & $0.10\pm0.03$~fb & 1.38~fb \\ 
		\hline
		\multicolumn{3}{c}{\emph{$\gamma\gamma+\MET$ with $gg\to H\to Sh\to \gamma\gamma\chi\chi$ at $\sqrt{s}=13$~TeV}}  \\
		\hline
		High $S_{\MET}$, high $p_\text{T}^{\gamma\gamma}$ & $0.265\pm0.009$~fb & 12.1~fb \\ 
		High $S_{\MET}$, low $p_\text{T}^{\gamma\gamma}$ & $0.675\pm0.014$~fb & 12.1~fb \\ 
		Intermediate $S_{\MET}$ & $3.17\pm0.03$~fb & 12.1~fb \\ 
		Rest & $2.80\pm0.03$~fb & 12.1~fb \\ 
		\hline
	\end{tabular}
	\caption{\label{tab:limits}Comparisons of the model's predictions for $gg\to H$ against (model-independent) visible cross section 95\% CLs in the CMS Run 1 monojet~\cite{Khachatryan:2014rra}, the ATLAS Run 2 $b\bar{b}+\MET$~\cite{ATLAS:2016tsc}, and the ATLAS Run 2 $\gamma\gamma+\MET$~\cite{ATLAS:2016thr} searches. For demonstration, the cross section of $gg\to H$ has been set to an optimistically high value of 10 (20)~pb for $\sqrt{s}=8~(13)$~TeV, and yet the prediction is still well within the limits. The mass and parameter points considered here correspond to those chosen in Sect.~\ref{sec:AZH}. Binomial errors on selection efficiencies have been incorporated into the theoretical predictions. The $\gamma\gamma+\MET$ experimental limit is not presented per category, so for each category the inclusive limit is shown.}
\end{table}

Using the model presented in this paper, a \textsc{Rivet} analysis was designed to mimic the ATLAS Run 2 $Z+\MET$ selection, and events were passed through this selection after being generated and showered at 13~TeV. The process which was generated is $gg\to A\to ZH$, and thereafter $Z\to\ell\ell$ (where $\ell=e,\mu$) and $H\to SS,Sh$. Both $S$ and $h$ are left open to decay, with $S$ at 150 GeV and having SM-like BRs as well as $BR(S\to\chi\chi)=0.5$. With $a_1=1$, the admixture of $SS$ and $Sh$ is considered to be equal. $m_H$ was considered at 300~GeV, $m_\chi=60$~GeV and $m_A$ took on the values 600 and 800~GeV. With this choice of parameters, the process described here is well within current limits for monojet and $b\bar{b}+\MET$ searches at the LHC, as discussed in Table~\ref{tab:limits}.

The results of this are shown in the first four plots in Fig.~\ref{fig:AZH}. Comparing with the distributions in Ref.~\cite{ATLAS_ZMET_RUN2}, the shapes of the distributions seem consistent with the data. The $p_\text{T}$ of the di-lepton system is sensitive to the mass of $A$, and can be used as a discriminant for its search. The selection efficiencies for the $m_A=600$ and 800~GeV simulations are 0.68\% and 1.86\% respectively. The ATLAS Run 2 excess of $\sim11$ events at $L=3.2~\text{fb}^{-1}$ can therefore be explained by a $gg\to A\to ZH$ production cross section in the order of tens of picobarns. However, contributions from $pp\to H\to SS,Sh$ production could also be a factor to account for, and in this case there would not only be contributions to the $Z$ peak region (i.e. where $m_{\ell\ell}\sim m_Z$), but also in the regions where $m_{\ell\ell}$ is significantly smaller or larger than $m_Z$. This is due to the fact that in $H\to SS,Sh$, $S$ can have a large BR to $WW$, and di-lepton pairs will come with \MET in the form of neutrinos for this decay, whereas jets could be found in the decay of the other $S$ or $h$.

The same events were passed through a selection mimicking the ATLAS Run 2 $A\to Zh$ (where $h\to b\bar{b}$) search~\cite{ATLAS_Zh_RUN2}. While there has so far been no significant excess in this channel, it is interesting to understand how the kinematics look for $A\to ZH$. The discriminant of these searches is typically the mass of the vector boson and Higgs boson pair, as reconstructed through a di-lepton and $b\bar{b}$ system in the 2 lepton category (for the 0 lepton category, a transverse mass is calculated instead). The mass of the $Zh$ system is shown by the last two plots in Fig.~\ref{fig:AZH}. On the right is the 1 $b$-tag category and on the left is the 2 $b$-tag category. Both plots are shown in the categories with low $p_\text{T}$ of the $Z$ (the high $p_\text{T}$ categories have a small selection efficiency). The selection efficiency is dominant in the 2 $b$-tag category with 2.2\% and 1.8\% for $m_A=600$ and 800~GeV respectively. The mass distributions do not peak at $m_A$ because the final state is not just $\ell\ell b\bar{b}$ -- more particles can come from the decay of $H\to SS,~Sh$, making the final state more diverse. Note that there is also a mass dependence on the $b$-tag categorisation. This is due to the fact that the $b\bar{b}$ system four vector is scaled to the Higgs mass in the analysis, whereas in this case $S\to b\bar{b}$ could also occur, distorting the kinematics.

\section{Summary}
\label{conc}

In this work we have presented the theory and rationale for introducing a number of new scalars to the SM. The particle content of the proposed model comes from a Type-II 2HDM and two new scalars, $S$ and $\chi$.

The study follows a previous work (in Ref.~\cite{vonBuddenbrock:2015ema}) which used $H$ and $\chi$ to predict a distorted Higgs boson $p_\text{T}$ spectrum through the effective decay $H\to h\chi\chi$. In this work, the effective interaction is assumed to be mediated by the scalar $S$, and $H$ is taken to be the heavy CP-even component of a Type-II 2HDM. The theoretical aspects of the equivalence between the effective model and the model presented in this paper is described in detail throughout Sects.~\ref{theory} and \ref{hpteff}.

With these new scalars, it is clear that a great deal of interesting phenomenology can be studied. Within certain mass ranges, a variety of signatures of the model have been discussed. $S$, in particular is a key element in the model, since it acts as a portal to DM interactions through its $S\to \chi\chi$ decay mode. It is also SM Higgs-like, and thus can be tagged through various decay modes. By a choice of parameters, it is assumed to be produced dominantly through the decay $H\to SS$ and $H\to Sh$, and is therefore likely to produce events that come with jets, leptons and \MET.

In addition to the discussion on the model, a few selected leptonic signatures have been explored using MC predictions and event selections. Various interesting distributions have been shown, as well as the rates and efficiencies of some processes which have relatively small SM backgrounds. The selected parameter points have also been compared to existing limits in the data, where applicable, and no violation of these limits has been found.

With the LHC continuing to deliver data at a staggering rate, it is important to keep testing models in the search for new physics. With a model dependence, experimentalists have a much clearer picture of what to look for in the data and how to bin results. It is evident that some hints exist in the search for new scalars at the LHC~\cite{vonBuddenbrock:2015ema}, and therefore the scalar sector is important to probe on both a theoretical and experimental level.

\begin{acknowledgements}
The work of N.C. and B. Mukhopadhyaya was partially supported by funding available from the Department of
Atomic Energy, Government of India for the Regional Centre for Accelerator-based Particle Physics (RECAPP), 
Harish-Chandra Research Institute. The Claude Leon Foundation are acknowledged for their financial support. 
The High Energy Physics group of the University of the Witwatersrand is grateful for the support from the Wits 
Research Office, the National Research Foundation, the National Institute of Theoretical Physics and the Department 
of Science and Technology through the SA-CERN consortium and other forms of support. T.M. is supported by funding 
from the Carl Trygger Foundation under contract CTS-14:206 and the Swedish Research Council under contract
621-2011-5107.
\end{acknowledgements}

\appendix

\section{Production: $gg \to A, H$}
\label{appenprod}
In Type-II 2HDMs, the $gg$F production cross section of the $CP$-odd Higgs $A$ is done by a 
simple rescaling of the SM Higgs ($h$) cross section~\cite{Coleppa:2013xfa}:
\begin{equation}
\sigma\left(gg \to h \right) \equiv \sigma_{\text{SM}}, 
\end{equation}
and is given as
\begin{equation}
\sigma\left(gg \to A \right) = \sigma_{\text{SM}} \times \frac{\left| \cot\beta F^A_{1/2}\left(\tau_t\right)
+ \tan\beta F^A_{1/2}\left(\tau_b\right)\right|^2}{\left|F^h_{1/2}\left(\tau_t\right) + F^h_{1/2}\left(\tau_b\right)\right|^2}.
\end{equation}
In this expression $\tau_f = 4 m_f^2/ m_A^2$ and the scalar and pseudoscalar loop factors are given by:
\begin{equation}
F^A_{1/2} = -2\tau f\left(\tau\right), \qquad\qquad F^h_{1/2} = -2 \tau \left[1+ \left(1-\tau\right) f\left(\tau\right)\right],
\label{loopf}
\end{equation}
where
\begin{equation}
f(\tau)
 =
 \begin{cases}
  \left[\sin^{-1} \left(1/\sqrt{\tau}\right)\right]^2 & \tau \geq 1, \\
  - \frac{1}{4} \left[ \ln\left(\frac{\eta_+}{\eta_-}\right) - i\pi\right]^2  & \tau < 1,
 \end{cases}
\end{equation}
with $\eta_{\pm} \equiv 1 \pm \sqrt{1-\tau}$. Here we have ignored the contributions of the other Higgs bosons in the loop, which
are typically small. Similarly, the $gg$F cross section for the $CP$-even Higgs (through a rescaling of the SM cross 
section) is given as:
\begin{equation}
\sigma\left(gg \to H \right) = \sigma_{\text{SM}} \times \frac{\left| \left(\frac{\sin\alpha}{\sin\beta} \right) F^h_{1/2}\left(\tau_t\right)
+ \left(\frac{\cos\alpha}{\cos\beta} \right) F^h_{1/2}\left(\tau_b\right)\right|^2}{\left|F^h_{1/2}\left(\tau_t\right) + F^h_{1/2}\left(\tau_b\right)\right|^2},
\end{equation}
where the loop factors (the $F$s) are defined in Eq.~\ref{loopf}.

\section{Interaction Lagrangians in 2HDM}
\label{intLag2hdm}
Interactions with electroweak vector bosons $V$ ($W^\pm, Z$) and the photon field ($A_\mu$) with $\phi$ and $H^\pm$
are given as:
\begin{align}
{\cal L}_{VV\phi} =&\, \frac{2 M_W^2}{v} \cos(\beta - \alpha) W^+_\mu W^{- \mu} H \notag \\
& + 2 \frac{M_W^2}{v} \left(\sin(\beta - \alpha) \right) W^+_\mu W^{- \mu} h \notag \\
& + \frac{M_Z^2}{v} \cos(\beta - \alpha) Z_\mu Z^\mu H
 + \frac{M_Z^2}{v} \left(\sin(\beta - \alpha)\right) Z_\mu Z^\mu h,
 \end{align}
 and,
 \begin{align}
{\cal L}_{V\phi\phi} =&\,  \frac{M_W}{v\,\cos\theta_W} \sin(\beta - \alpha) Z_\mu
\left(A\partial_\mu H - H\partial_\mu A \right) \notag \\
 &+ \frac{M_W}{v\,\cos\theta_W} \cos(\beta - \alpha) Z_\mu (A\partial_\mu h - h\partial_\mu A) \notag \\
& + i \frac{M_W}{v} \frac{(2\,\cos^2\theta_W - 1)}{\cos\theta_W} Z_\mu \left(H^-\partial_\mu H^+  - H^+\partial_\mu H^- \right) \notag\\
& + i e A_\mu \left(H^-\partial_\mu H^+  - H^+\partial_\mu H^- \right) \notag \\
& + \Bigg[ i \frac{M_W}{v} \sin(\beta - \alpha) \left(W^{-\mu} H\partial_\mu H^+  - W^{-\mu} H^+ \partial_\mu H \right) 
 \notag \\
& +  i \frac{M_W}{v} \cos(\beta - \alpha) \left( W^{-\mu} h\partial_\mu H^+  - W^{-\mu} H^+ \partial_\mu h \right) 
 \notag \\
& + \frac{M_W}{v} \left(W^{-\mu} A\partial_\mu H^+  - W^{-\mu} H^+ \partial_\mu A \right) + \text{h.c} \Bigg]. \label{eqn:vphiphi}
\end{align}
In a Type-II 2HDM framework, the Yukawa terms are as follows:
\begin{align}
{\cal L}^{Y}_{h} =&\,  -\frac{1}{v}\left[ \frac{\cos\alpha}{\sin\beta} \sum_{q_u} y_{m_{q_u}} q_u \bar q_u h 
+\frac{\sin\alpha}{\cos\beta} \sum_{q_d} y_{m_{q_d}} q_d \bar q_d h \right],\\
{\cal L}_{H}^{Y} =&\,  -\frac{1}{v}\left[\frac{\sin\alpha}{\sin\beta} \sum_{q_u} y_{m_{q_u}} q_u \bar q_u H
+ \frac{\cos\alpha}{\cos\beta} \sum_{q_d} y_{m_{q_d}} q_d \bar q_d H \right], \\
{\cal L}_{A}^{Y} =&\,  - \frac{i}{v} \left[ \cot\beta
  \sum_{q_u} y_{m_{q_u}} q_u \gamma_5 \bar q_u A + \tan\beta \sum_{q_d} y_{m_{q_d}} q_d \gamma_5 \bar q_d A \right],  \\
{\cal L}_{H^\pm}^{Y} =&\,  \frac{1}{2} \Big[ \left(-y_{ut} \cos\beta + y_{ub} \sin\beta\right) \left(\bar t b H^+ + \bar b t H^-\right) \notag \\
&\,\,\,\,+   \left(y_{ut} \cos\beta + y_{ub} \sin\beta\right) \left(\bar t \gamma_5 b H^+ - \bar b \gamma_5 t H^-\right)\Big],
\end{align}
with $y_{ut} = \sqrt{2} y_{m_t}/(v \sin\beta)$ and $y_{ub} = \sqrt{2} y_{m_b}/(v \cos\beta)$. 
The relevant trilinear scalar interactions are part of the Lagrangian ${\cal L}_{\phi\phi\phi}$, 
\begin{align}
{\cal L}_{\phi\phi\phi} =&\,  - v \lambda_{h H^+ H^-} h H^+ H^- - v \lambda_{h H^+ H^-} H H^+ H^-  \notag\\
& - \frac{1}{2} v \lambda_{H h h} H h^2,
\end{align}
where the couplings have the following expressions:
\begin{align}
\lambda_{h H^+ H^-} =&\, \frac{-1}{2 v^2 \sin(2\beta)} \Big[  m_h^2 \cos(\alpha-3 \beta) + 3 m_h^2 \cos(\alpha+\beta) \notag\\
&\, -4 m_{H^\pm}^2 \sin(2\beta) \sin(\alpha-\beta) - 4 M^2 \cos(\alpha+\beta)   \Big], \\
\lambda_{H H^+ H^-} =&\, \frac{-1}{2 v^2 \sin(2\beta)} \Big[  m_H^2 \sin(\alpha-3 \beta) + 3 m_h^2 \sin(\alpha+\beta) \notag\\
&\, + 4 m_{H^\pm}^2 \sin(2\beta) \cos(\alpha-\beta)  - 4 M^2 \sin(\alpha+\beta)   \Big], \\
\lambda_{H h h} =& \frac{-1}{2 v^2 \sin(2\beta)} \Big[  (2 m_h^2 + m_H^2) \cos(\alpha-\beta)\sin(2 \alpha) \notag \\
&\, - M^2 \cos(\alpha-\beta) (3 \sin(2\alpha) - \sin(2\beta))
   \Big].
\end{align}
Here $M^2$ is the shorthand notation for $m^2_{12}/(\sin\beta \cos\beta)$.

\end{document}